\documentclass[longauth]{aa}  

\usepackage{graphicx}
\usepackage{txfonts}
\begin{document}

   \title{TOI-674b: an oasis in the desert of exo-Neptunes transiting a nearby M dwarf
   \thanks{Based on observations made with the HARPS instrument on the ESO 3.6-m telescope at La Silla Observatory under programme ID 1102.C-0339} }
   
   \author{F. Murgas\inst{1,2}
          \and
          N. Astudillo-Defru\inst{3}
          \and
          X. Bonfils\inst{4}
          \and
          Ian Crossfield\inst{5}
          \and
          J. M. Almenara\inst{4}
          \and
          John Livingston\inst{6}
          \and
          Keivan G. Stassun\inst{7}
          \and
          Judith Korth\inst{35}
          \and
          Jaume Orell-Miquel\inst{1,2}
          \and
          G. Morello\inst{1,2}
          \and
          Jason D. Eastman\inst{14}
          \and
          Jack J. Lissauer\inst{9}
          \and
          Stephen R. Kane\inst{32}
          \and
          Farisa Y. Morales\inst{33}
          \and
          Michael W. Werner\inst{33}
          \and
          Varoujan Gorjian\inst{33}
          \and
          Bj\"{o}rn Benneke\inst{22}
          \and
          Diana Dragomir\inst{34}
          \and
          Elisabeth C. Matthews\inst{8}
          \and
          Steve B. Howell\inst{9}
          \and
          David Ciardi\inst{10}
          \and
          Erica Gonzales\inst{11}
          \and
          Rachel Matson\inst{12}
          \and
          Charles Beichman\inst{10}
          \and
          Joshua Schlieder\inst{13}
          \and
          Karen A.\ Collins\inst{14}
          \and
          Kevin I.\ Collins\inst{15}
          \and
          Eric L.\ N.\ Jensen\inst{16}
          \and
          Phil Evans\inst{17}
          \and
          Francisco J. Pozuelos\inst{18,19}
          \and
          Micha\"el Gillon\inst{19}
          \and
          Emmanu\"el Jehin\inst{18}
          \and
          Khalid Barkaoui\inst{19,20}
          \and
          E. Artigau\inst{22}
          \and
          F. Bouchy\inst{8}
          \and
          D. Charbonneau\inst{14}
          \and
          X. Delfosse\inst{4}
          \and
          R. F. D\'{i}az\inst{21}
          \and
          R. Doyon\inst{22}
          \and
          P. Figueira\inst{23,24}
          \and
          T. Forveille\inst{4}
          \and
          C. Lovis\inst{8}
          \and
          C. Melo\inst{23}
          \and
          G. Gaisn\'{e}\inst{4}
          \and
          F. Pepe\inst{8}
          \and
          N. C. Santos\inst{24,25}
          \and
          D. S\'{e}gransan\inst{8}
          \and
          S. Udry\inst{8}
          \and
          Robert~F.~Goeke\inst{28}
          \and
          Alan~M.~Levine\inst{26}
          \and
          Elisa~V.~Quintana\inst{13}
          \and
          Natalia M. Guerrero\inst{26}
          \and 
          Ismael Mireles\inst{26}
          \and
          Douglas A. Caldwell\inst{30,9}
          \and
          Peter Tenenbaum\inst{30,9}
          \and
          C. E. Brasseur\inst{31}
          \and
          G. Ricker\inst{26}
          \and
          R. Vanderspek\inst{26}
          \and
          David W. Latham\inst{14}
          \and
          S. Seager\inst{26,27,28}
          \and
          J. Winn\inst{29}
          \and
          Jon M. Jenkins\inst{9}
          }

   \institute{Instituto de Astrof\'{i}sica de Canarias (IAC), 38205 La Laguna, Tenerife, Spain\\
        \email{fmurgas@iac.es}
        \and
        Departamento de Astrof\'{i}sica, Universidad de La Laguna (ULL), 38206 La Laguna, Tenerife, Spain
        \and
        Departamento de Matem\'{a}tica y F\'{i}sica Aplicadas, Universidad Cat\'{o}lica de la Sant\'{i}sima Concepci\'{o}n, Alonso de Rivera 2850, Concepci\'{o}n, Chile
        \and
        Univ. Grenoble Alpes, CNRS, IPAG, F-38000 Grenoble, France
        \and
        Department of Physics and Astronomy, University of Kansas, Lawrence, KS, USA
        \and
        Department of Astronomy, University of Tokyo 7-3-1 Hongo, Bunkyo-ku, Tokyo 113-0033, Japan
        \and
        Vanderbilt University, Department of Physics \& Astronomy, 6301 Stevenson Center Ln., Nashville, TN 37235, USA
        \and
        Observatoire de l’Universit\'e de Gen\`eve, Chemin des Maillettes 51, 1290 Versoix, Switzerland
        \and
        NASA Ames Research Center, Moffett Field, CA, 94035, USA
        \and
        Caltech/IPAC, 1200 E. California Blvd. Pasadena, CA 91125, USA
        \and
        Department of Astronomy and Astrophysics, University of California, Santa Cruz, CA 95064, USA
        \and
        U.S. Naval Observatory, Washington, DC 20392, USA
        \and
        NASA Goddard Space Flight Center, 8800 Greenbelt Rd, Greenbelt, MD 20771, USA
        \and
        Center for Astrophysics \textbar \ Harvard \& Smithsonian, 60 Garden Street, Cambridge, MA 02138, USA
        \and
        George Mason University, 4400 University Drive, Fairfax, VA, 22030 USA
        \and
        Department of Physics \& Astronomy, Swarthmore College, Swarthmore PA 19081, USA
        \and
        El Sauce Observatory, Coquimbo Province, Chile
        \and
        Space Sciences, Technologies and Astrophysics Research (STAR) Institute, Universit\'e de Li\`ege, 19C All\`ee du 6 Ao\^ut, 4000 Li\`ege, Belgium
        \and
        Astrobiology Research Unit, Universit\'e de Li\`ege, 19C All\`ee du 6 Ao\^ut, 4000 Li\`ege, Belgium
        \and
        Oukaimeden Observatory, High Energy Physics and Astrophysics Laboratory, Cadi Ayyad University, Marrakech, Morocco
        \and
        International Center for Advanced Studies (ICAS) and ICIFI(CONICET), ECyT-UNSAM, Campus Miguelete, 25 de Mayo y Francia (1650), Buenos Aires, Argentina
        \and
        Universit\'{e} de Montr\'eal, D\'epartement de Physique \& Institut de Recherche sur les Exoplan\`etes, Montr\'eal, QC H3C 3J7, Canada
        \and
        European Southern Observatory, Alonso de C\'{o}rdova 3107, Vitacura, Regi\'on Metropolitana, Chile
        \and
        Instituto de Astrof\'{i}sica e Ci\^{e}ncias do Espa\c{c}o, Universidade do Porto, CAUP, Rua das Estrelas, 4150-762 Porto, Portugal
        \and
        Departamento de F\'{i}sica e Astronomia, Faculdade de Ci\^{e}ncias, Universidade do Porto, Rua do Campo Alegre, 4169-007 Porto, Portugal
        \and
        Department of Physics and Kavli Institute for Astrophysics and Space Research, Massachusetts Institute of Technology, Cambridge, MA 02139, USA
        \and
        Department of Earth, Atmospheric and Planetary Sciences, Massachusetts Institute of Technology, Cambridge, MA 02139, USA
        \and
        Department of Aeronautics and Astronautics, MIT, 77 Massachusetts Avenue, Cambridge, MA 02139, USA
        \and
        Department of Astrophysical Sciences, Princeton University, NJ 08544, USA
        \and
        SETI Institute, 189 Bernardo Ave, Suite 200, Mountain View, CA 94043, USA
        \and
        Space Telescope Science Institute, Baltimore, MD 21218, USA
        \and
        Department of Earth and Planetary Sciences, University of California, Riverside, CA 92521, USA
        \and
        Jet Propulsion Laboratory, California Institute of Technology, 4800 Oak Grove Drive, Pasadena, CA 91109, USA
        \and
        Department of Physics and Astronomy, University of New Mexico, 210 Yale Blvd NE, Albuquerque, NM 87106, USA
        \and
        Department of Space, Earth and Environment, Astronomy and Plasma Physics, Chalmers University of Technology, SE-412 96 Gothenburg, Sweden
        }

   \date{Received 2021; accepted 2021}

 
  \abstract
  {The NASA mission \textit{TESS} is currently doing an all-sky survey from space to detect transiting planets around bright stars. As part of the validation process, the most promising planet candidates need to be confirmed and characterised using follow-up observations.}
   {In this article we aim to confirm the planetary nature of the transiting planet candidate TOI-674b using spectroscopic and photometric observations.}
   {We use \textit{TESS}, \textit{Spitzer}, ground-based light curves and HARPS spectrograph radial velocity measurements to establish the physical properties of the transiting exoplanet candidate TOI-674b. We perform a joint fit of the light curves and radial velocity time series to measure the mass, radius, and orbital parameters of the candidate.}
   {We confirm and characterize TOI-674b, a low-density super-Neptune transiting a nearby M dwarf. The host star (TIC 158588995, $V = 14.2$ mag, $J = 10.3$ mag) is characterized by its M2V spectral type with $\mathrm{M}_\star=0.420\pm 0.010$ M$_\odot$, $\mathrm{R}_\star = 0.420\pm 0.013$ R$_\odot$, and $\mathrm{T}_{\mathrm{eff}} = 3514\pm 57$ K, and is located at a distance $d=46.16 \pm 0.03$ pc. Combining the available transit light curves plus radial velocity measurements and jointly fitting a circular orbit model, we find an orbital period of $1.977143 \pm 3\times 10^{-6}$ days, a planetary radius of $5.25 \pm 0.17$ $\mathrm{R}_\oplus$, and a mass of $23.6 \pm 3.3$ $\mathrm{M}_\oplus$ implying a mean density of $\rho_\mathrm{p} = 0.91 \pm 0.15$ [g cm$^{-3}$]. A non-circular orbit model fit delivers similar planetary mass and radius values within the uncertainties. Given the measured planetary radius and mass, TOI-674b is one of the largest and most massive super-Neptune class planets discovered around an M type star to date. It is also a resident of the so-called Neptunian desert and a promising candidate for atmospheric characterisation using the James Webb Space Telescope.}
   {}

   \keywords{stars: individual: TOI-674 -- planetary systems -- techniques: photometric -- techniques: radial velocities}

   \maketitle
%

\section{Introduction}

M dwarf stars are the most common type of stars in the Milky Way and in the solar neighborhood (e.g., \citealp{Reid1995}, \citealp{Chabrier2000}, \citealp{Henry2006}, \citealp{Bochanski2010}, \citealp{Winters2015}). Due to their relatively small sizes and low masses, they are good targets for the detection of small planets using transit searches and radial velocity measurements, respectively. More importantly, transiting planets orbiting bright M dwarfs offer great opportunities for atmospheric characterization using, for example, low and high resolution transmission spectroscopy (e.g., \citealp{Kreidberg2014}, \citealp{Knutson2014}, \citealp{Ehrenreich2014}, \citealp{Southworth2017}). Several Earth-sized and super-Earth planets have been found orbiting around M dwarfs (e.g., GJ 876d \citealp{Rivera2005}, TRAPPIST-1 system \citealp{Gillon2017}, LHS 1140b \citealp{Dittmann2017}, LHS 1140c \citealp{Ment2019}, GJ 357b \citealp{Luque2019}), but also planets with masses and radii between those of Neptune and Jupiter have been found (e.g., \hbox{GJ 876c} \citealp{Marcy2001}, \hbox{GJ 436b} \citealp{Gillon2007}; \hbox{HATS-71b} \citealp{Bakos2020}; \hbox{GJ 3512b} \citealp{Morales2019}; \hbox{TOI-1728b} \citealp{Kanodia2020}). Relatively few gas planets orbiting around M dwarfs have been found in this mass and size range, in agreement with the prediction of a paucity of gas giants orbiting around M dwarfs for core accretion models (\citealp{Laughlin2004}). Specifying their occurrence rate with better statistics can give new insights on planetary formation (e.g., core accretion versus disk instability: \citealp{Boss2006}) and orbital migration processes (e.g., \citealp{Correia2020}).

Planetary population studies made for different types of stars have noticed a lack of Neptune-sized planets with short orbital periods (i.e., highly irradiated), this is the so-called `Neptunian desert' (e.g., \citealp{Szabo2011}, \citealp{Mazeh2016}, \citealp{Fulton&Petigura2018}). The low number of Neptune-sized planets with orbital periods shorter than 4 days could possibly indicate different formation mechanisms for close-in super-Earths and Jovian planets (\citealp{Mazeh2016}), similar to what is observed for low-mass brown dwarfs in short orbits around Sun-like stars (e.g., \citealp{Grether2006}). Another proposed mechanism for the origin of this gap in the radius distribution is photo-evaporation of planetary atmospheres in response to high energy radiation (ultraviolet, X-ray) from the star (e.g., \citealp{Lopez2013}, \citealp{Chen2016}).

In recent years, more planets in the Neptunian desert have been identified. Some examples of planets with masses close to Neptune and in short orbital periods found recently are NGTS-4b (\citealp{West2019}), TOI-132b (\citealp{Diaz2020}), LTT 9779b (\citealp{Jenkins2020}), TOI-1728b (\citealp{Kanodia2020}), TOI-442b (\citealp{Dreizler2020}), TOI-849b (\citealp{Armstrong2020}). The addition of new objects in this mass, radius, and orbital period range can help to shed some light on the physical mechanisms behind the Neptunian desert.

The Transiting Exoplanet Survey Satellite (\textit{TESS}; \citealp{Ricker2014}) is a NASA-sponsored space telescope launched in April 2018. Its main mission is to monitor the full sky in search of transiting planets orbiting around bright stars ($5 < I_C < 13$ mag). The observing strategy of \textit{TESS} is to observe with its 4 cameras a $24^\circ \times 96^\circ$ area of the sky for 27 days. The exposure time for each camera is of 2 seconds and during the Primary Mission the images were stacked in two timing sampling modes: 2 minute and 30 minute cadence. The planned time for the main mission was 2 years, but on July 2020 \textit{TESS} entered its extended mission which is approved to continue until the end of September 2022. For the first \textit{TESS} extended mission the 30-minute cadence for Full Frame Images (FFIs) has been reduced to 10 minutes.

Here, we report the discovery of TOI-674b, a super-Neptune transiting around a nearby M dwarf, discovered using \textit{TESS} data. The mass, radius, and orbital period of this planet indicate that this is a new member of the Neptunian desert, and it is a good candidate target for follow-up observations. This paper is organized as follows: Section \ref{Sec:Obs} presents the observations used in this work, Section \ref{Sec:Analysis} presents the methods use in the data analysis, Section \ref{Sec:Results} presents the parameter estimates of the planet as well as a discussion in the context of known planets. Finally, Section \ref{Sec:Conclusions} presents the conclusions of this work.


\section{Observations}
\label{Sec:Obs}

\subsection{Space-based photometry}
\subsubsection{\textit{TESS} photometry}
The star TIC 158588995 (TOI-674) was observed by \textit{TESS} during Sector 9 and Sector 10 for 27 days each. The Sector 9 campaign started on 28 February 2019 and ended on 25 March 2019; for this campaign the target was positioned on \textit{TESS} CCD 3 Camera 2. The Sector 10 campaign started on 26 March 2019 and ended on 22 April 2019, and for this \textit{TESS} run the target was placed on CCD 4 Camera 2. For its prime mission, \textit{TESS} images were stacked in two timing sampling modes: 2 minute and 30 minute cadence; TOI-674 was selected to be observed using the 2 minute short-cadence mode in both sectors (\citealp{Stassun2018}).

The raw image data taken by \textit{TESS} was processed by the Science Processing Operations Center (SPOC) at NASA Ames Research Center. The SPOC pipeline (\citealp{Jenkins2016}) calibrates the image data, performs quality control (e.g., identifies and flags bad data), extracts photometry for each target star in the \textit{TESS} field of view, and searches the resulting light curves for exoplanet transit signatures. The initial TESS light curve was produced using simple aperture photometry (SAP, \citealp{Morris2020}) and then the instrumental systematic effects were removed using the Presearch Data Conditioning (PDC) pipeline module (\citealp{Smith2012}, \citealp{Stumpe2014}). The light curves are searched with an adaptive, wavelet-based matched filter (\citealp{Jenkins2002}, \citealp{JenkinsJM2020}) , and then fitted to limb-darkened transit models (\citealp{Li2019}) , and subjected to diagnostic tests to make or break the planetary hypothesis (\citealp{Twicken2018}). Subsequently, the \textit{TESS} science office reviewed the transit signature identified in the SPOC processing of Sector 9, promoting it to a planetary candidate as TESS Object of Interest (TOI) TOI-674.01, and alerted the community in May 2019. 

\begin{figure*}
   \centering
   \includegraphics[width=\hsize]{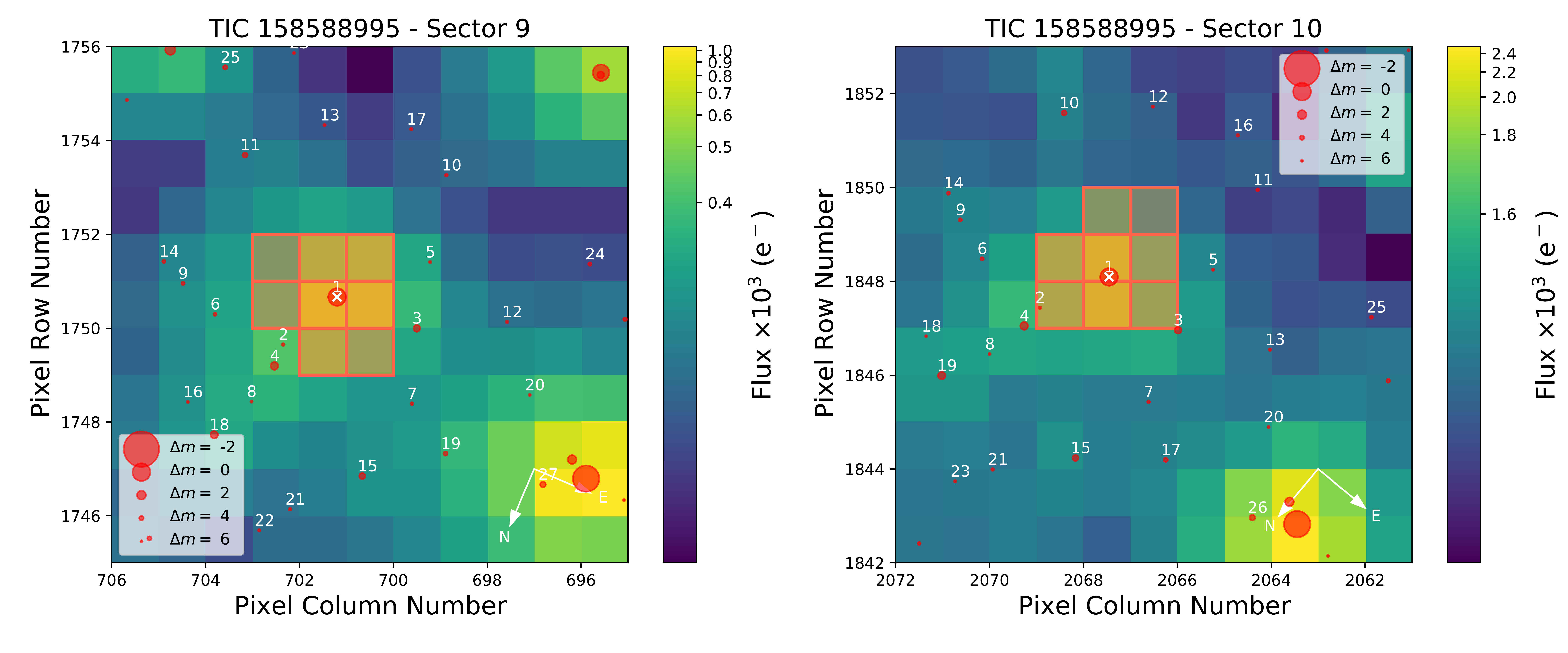}
   \caption{\textit{TESS} target pixel file image of TOI 674 observed in Sector 9 (left) and Sector 10 (right), made with \texttt{tpfplotter} (\citealp{Aller2020}). The pixels highlighted in red show the aperture used as the target region in the photometry. The position and sizes of the red circles represent the position and \textit{TESS} magnitudes of nearby stars respectively.}
   \label{Fig:TESS_TPF_S09andS10}
\end{figure*}

In this work we made used of the \textit{TESS} PDC- SAP light curves that have been corrected for instrumental systematics. These data sets (TESS sector 9 and 10) are available at the Barbara A. Mikulski Archive for Space Telescopes (MAST\footnote{\url{https://mast.stsci.edu/portal/Mashup/Clients/Mast/Portal.html}}). The \textit{TESS} images around the position of TOI-674 in Sector 9 and 10 are shown in Figure \ref{Fig:TESS_TPF_S09andS10}. The red squares indicate the aperture used to produce the PDC-SAP light curves.

\begin{table*}
  \caption[]{TOI-674 identifiers, coordinates, stellar parameters, and magnitudes. The adopted stellar mass and radius used in this work are the ones presented in the middle column (i.e., BT-Settl models SED fit).}
  \label{Table:TOI674dinfo}
  \centering
  \begin{tabular}{llcccc}
    \hline \hline
    Identifiers & & & & & Ref. \\
    \hline
    TIC & & & & 158588995 & \\
    2MASS & & & & J10582099-3651292 & \\
    Gaia EDR3 & & & & 5400949450924312576 & \\
    \hline
    Equatorial coordinates & & & & & \\
    \hline
    RA (J2016) & & & & $10^{\mathrm{h}}\; 58^{\mathrm{m}}\; 20^{\mathrm{s}}.77$ & 1 \\
    DEC (J2016) & & & & $-36^{\circ}\; 51'\; 29\farcs19 $ &  1 \\ 
    $\mu_{\alpha}$ [mas/year] & & & & $-163.255 \pm 0.015$ & 1 \\
    $\mu_{\delta}$ [mas/year] & & & & $-3.672 \pm 0.015$ & 1 \\
    Parallax [mas] & & & & $21.623 \pm 0.016$ & 1 \\
    \hline
    Parameters from SED fit & & BT-Settl SED fit &  & BT-Settl + Stellar evol. models SED fit & \\
    \hline
    Effective Temperature [K]\tablefootmark{a} & $\mathrm{T}_{\mathrm{eff}}$ & $3514\pm 57$ &  & $3505^{+28}_{-32}$ &  5 \\
    Stellar Luminosity [$\mathrm{L}_\odot$] & $\mathrm{L}$ & $0.0243^{+0.0012}_{-0.0009}$ &  & $0.0240^{+0.0013}_{-0.0008}$ & 5 \\
    Surface gravity [cm/s$^2$]\tablefootmark{b} & $\log(g)$ & $5.28^{+0.51}_{-0.66}$ &  & $4.832^{+0.006}_{-0.009}$ & 5 \\
    Metallicity [dex]\tablefootmark{a} & $[\mathrm{Fe}/\mathrm{H}]$ & $0.17\pm 0.12$ &  & $0.114\pm 0.074$ & 5 \\
    Stellar Age [Gyr] & & & & $5.5^{+2.9}_{-1.9}$ &  5 \\
    Mass   [M$_\odot$] & $\mathrm{M}_\star $ & $0.420\pm 0.010$\tablefootmark{c} &  & $ 0.442 \pm 0.007 $ & 5 \\
    Radius [R$_\odot$] & $\mathrm{R}_\star $ & $0.420\pm 0.013$ &  & $ 0.421 \pm 0.007 $  & 5 \\
    Stellar density [g cm$^{-3}$] & $\rho_\star$ & $7.99 \pm 0.76$ &  & $8.36^{+0.22}_{-0.31} $ & 5 \\
    Surface gravity [cm/s$^2$] & $\log(g)$ & $4.81 \pm 0.03$\tablefootmark{d} &  & $4.832^{+0.006}_{-0.009}$ & 5 \\
    Distance [pc] & $\mathrm{d}_\star$ & $46.16\pm 0.03$ &  & $46.16\pm 0.03$ & 5 \\
    \hline
    Other parameters & & & & & \\
    \hline
    Stellar activity index & $\log_{R_{HK}}$ &  & $-5.14 \pm 0.01$ &  & 5 \\
    Stellar rotation period [days]\tablefootmark{e} & $P_{\mathrm{rot}}$ & & $52 \pm 5$ & & 5 \\
    Stellar rotation velocity [km/s]\tablefootmark{f} & &  & $0.4 \pm 0.04$ &  & 5 \\
    \hline
    Apparent magnitudes & & & & & \\
    \hline
    Gaia G [mag] & & & & $13.073\pm 0.002$ & 1 \\
    B [mag] & & & & $15.742\pm 0.024$ & 2 \\
    V [mag] & & & & $14.203\pm 0.041$ & 2 \\
    Sloan g [mag] & & & & $14.938\pm 0.022$ & 2 \\
    Sloan r [mag] & & & & $13.626\pm 0.047$ & 2 \\
    Sloan i [mag] & & & & $12.299\pm 0.068$ & 2 \\
    J [mag] & & & & $10.359\pm 0.023$ & 3 \\
    H [mag] & & & & $9.737\pm 0.023$ & 3 \\
    K [mag] & & & & $9.469\pm 0.019$ & 3 \\
    WISE W1 [mag] & & & & $9.345\pm 0.022$ & 4 \\
    WISE W2 [mag] & & & & $9.237\pm 0.019$ & 4 \\
    WISE W3 [mag] & & & & $9.163\pm 0.038$ & 4 \\
    WISE W4 [mag] & & & & $8.332\pm 0.455$ & 4 \\
    \hline
  \end{tabular}
  
  \tablebib{ (1)~\citet{GaiaEDR32020}; (2) \citet{Henden2015}; (3) \citet{Cutri2003}; (4) \citet{Cutri2014}; (5) This work.
  }
  
  \tablefoot{
  \tablefoottext{a}{As an independent check, we also derived Teff and [Fe/H] using the Machine Learning approach implemented in ODUSSEAS (\citealp{Karnavas2020}). The obtained Teff (3220$\pm$70 K) is lower than the one obtained with \texttt{SpecMatch-Emp} (\citealp{Yee2017}), while the [Fe/H] (+0.05$\pm$0.11) is compatible within the 1-$\sigma$ uncertainties.}
  \tablefoottext{b}{Stellar surface gravity $\log(g)$ derived from SED fit.}
  \tablefoottext{c}{Mass derived using the empirical relations of \citet{Mann2019}.}
  \tablefoottext{d}{Stellar surface gravity $\log(g)$ computed using the derived stellar mass and radius.}
  \tablefoottext{e}{From empirical relations of \citet{Astudillo-Defru:2017}.}
  \tablefoottext{f}{From stellar radius and rotation period.}
  }
\end{table*}

\subsubsection{\textit{Spitzer} photometry}
A primary transit of TOI-674b was observed using the \textit{Spitzer} Space Telescope as part of a program dedicated to \textit{TESS} target follow-up (GO-14084, \citealp{Crossfield2018}). The data were taken with the instrument InfraRed Array Camera (IRAC), which is a four-channel camera with a field of view of $5.2' \times 5.2'$ and can take images at 3.6, 4.5, 5.8, and 8 $\mu$m. Each channel has $256 \times 256$ pixels and a pixel scale of $\sim 1.2$ \arcsec{} pixel$^{-1}$. The target was observed on 29 September 2019 using \textit{Spitzer}'s 4.5 $\mu$m channel, the exposure time was set to 2 seconds and the total observation time was $\sim 5$ hours. To compute the time series we used a $3\times 3$ pixel subarray centered on the position of TOI-674.

\subsection{Ground-based photometry}
After the discovery of the transits made by the \textit{TESS} team, several follow-up ground-based observations were scheduled to confirm that the transits occur on TOI-674 and rule-out some false positive scenarios (e.g., eclipsing binaries unresolved by \textit{TESS}). The data were acquired under the \textit{TESS} Follow-up Observing Program (TFOP) and uploaded to the Exoplanet Follow-up Observing Program for \textit{TESS} (ExoFOP-TESS\footnote{\url{https://exofop.ipac.caltech.edu/tess/}}). 

\subsubsection{El Sauce Observatory}
We observed a full transit of TOI-674b in the Rc band on 12 May 2019 from El Sauce Observatory in Coquimbo Province, Chile. The 0.36 m telescope is equipped with a $1536\times1024$ SBIG STT-1603-3 camera. The camera was operated using a $2\times 2$ bin mode, the image scale is 1.47 \arcsec{} pixel$^{-1}$ resulting in a $18.8\arcmin\times12.5\arcmin$ field of view. The photometric data were extracted using \texttt{AstroImageJ} (\citealt{Collins2017}) with a 5 pixel aperture.

\subsubsection{LCOGT}
Las Cumbres Observatory Global Telescope Network (LCOGT, \citealp{Brown2013}) is a set of robotic telescopes operating in both hemispheres and with sites distributed across several countries. A primary transit of TOI-674b was observed on 16 May 2019 with the LCOGT 1 m telescope at Cerro Tololo International Observatory (CTIO) in Chile. The LCOGT 1 m telescopes are equipped with SINISTRO CCDs with a field of view of $26' \times 26'$ , and feature a pixel scale of 0.389 arcsec/pixel and read out cadence of 28 seconds. The transit was observed in the Sloan g band, the telescope was slightly defocused (0.1 mm from nominal value), and the exposure time was set to 150 seconds. The photometry was extracted using \texttt{AstroImageJ} software  with an aperture of 10 pixels and the Point Spread Function (PSF) diameter was 1.95~arcsec. 

\subsubsection{TRAPPIST-South}
TRAPPIST-South, located at ESO-La Silla Observatory in Chile, is a 60 cm Ritchey–Chretien telescope equipped with a thermoelectrically cooled $2K\times2K$ FLI Proline CCD camera (\citealp{Jehin2011,gillon2013}). It features a field- of-view of $22'\times22'$ and a pixel-scale of 0.65 \arcsec{}/pixel. We acquired 494 images during a full-transit observation on 2019-05-13 with the $I+z$ filter and an exposure time of 15 s. The optimum photometric aperture was 6 pixels (3.9 arcsec) and the PSF diameter 2.2~arcsec. A second full-transit observation was carried out on 2019-05-21 using the Sloan z filter with an exposure time of 15 s, and yielded a total of 412 images. The optimum aperture was 8~pixels (5.2~arcsec) and the PSF diameter was 2.5~arcsec. These sets of observations confirmed the presence of a transit event on the target star on time, and eliminated the possibility that the events are due to an eclipsing binary outside of the PSF centered on the target star. For both dates we made use of the \textit{TESS} Transit Finder tool, which is a customized version of the \texttt{Tapir} software package \citep{jensen2013}, to schedule the photometric time-series and we used \texttt{AstroImageJ} to perform aperture photometry.

\subsection{Spectroscopic observations}
TOI-674 was observed by the High Accuracy Radial velocity Planet Searcher (HARPS, \citealp{Mayor2003}) at the ESO La Silla 3.6m telescope. The observations were carried out as part of the program 1102.C-0339 dedicated to searching for planets orbiting around M dwarfs. From 24 May 2019 to 18 July 2019, we collected 17 spectra with a resolving power of $R \approx115000$. The image of the target star was placed in the aperture of the scientific fiber while the calibration fiber was used to monitor the sky, the exposure time was set to 30 minutes and the CCD was read using the slow read-out mode, resulting in a signal-to-noise ratio between 5 and 13 (median of 9).

The spectra were calibrated and extracted using the HARPS online pipeline (\citealp{Lovis2007}; for improvements in the pipeline check \citealp{Mayor2009b,Mayor2009a} and references therein). With the spectra and preliminary radial velocity measurements from the HARPS online pipeline we produced a template spectrum following \cite{Astudillo2017}. The final radial velocities were obtained from likelihood functions constructed by comparing the stellar template shifted by different velocities with each individual spectrum. The velocity that produced the maximum likelihood for each data point was the derived HARPS radial velocity. The RV time series present a dispersion of 17.4 m/s and have a median uncertainty of 7.3 m/s.

We also obtain from the spectra several activity indices: $\mathrm{H}_\alpha$, $\mathrm{H}_\beta$, $\mathrm{H}_\gamma$, Na, and calcium S index. The radial velocities and activity indices are given in Table \ref{Table:HARPS_RV}. 

\subsection{High-resolution imaging}

\textit{TESS} has a pixel scale of 21 arcsec/pixel, hence it is possible that TOI-674 can present some level of flux contamination produced by nearby faint stars not detected by seeing limited photometric observations. Contamination in the flux of the transit host star can led to a wrong estimation of transit depth, thus leading to an incorrect absolute planet radius. \textit{TESS} Data Validation Report (\citealp{Twicken2018}) performs a difference image centroiding analysis, in the case of TOI-674 the report determined that the location of the source of the transit signature was within 1 arcsec of the target star. Nevertheless, we searched for previously undetected faint nearby companion stars using adaptive optics and speckle imaging using Gemini Telescope.

\subsubsection{Gemini/NIRI}
\begin{figure}
   \centering
   \includegraphics[width=\hsize]{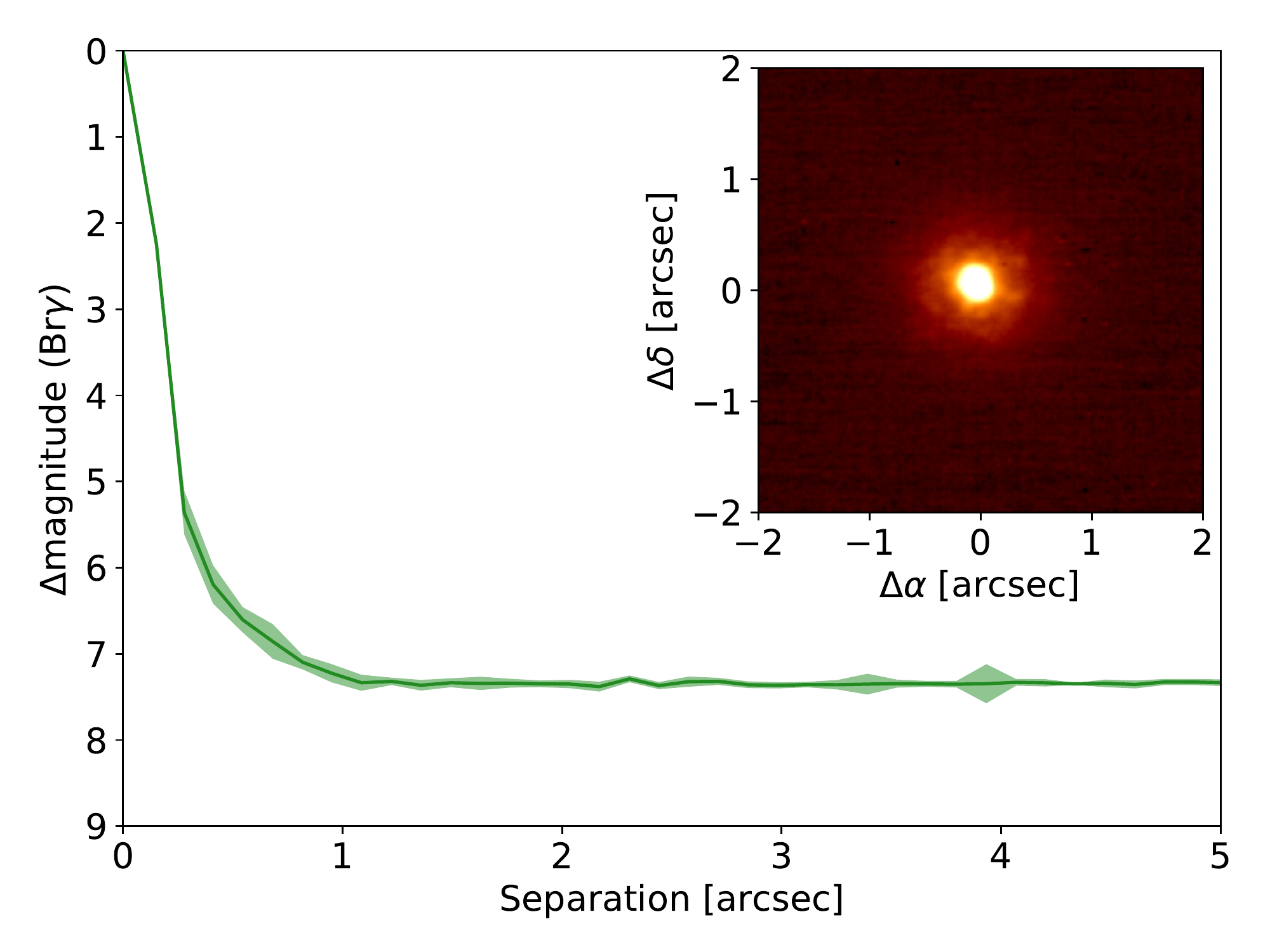}
   \caption{Gemini/NIRI high resolution image of TOI-674. Our observations rule out candidates to 5 mag fainter than the host beyond 270 mas, and 7.3 mag beyond $\sim$1.05\arcsec{}. \textit{Inset:} thumbnail image of TOI-674, centered on the star. The PSF is circular to the limit of our resolution, and no visual companions are identified anywhere in the field of view, which extends at least 13\arcsec{} from the target in all directions.}
   \label{Fig:AO_NIRI}
 \end{figure}

We searched for visual companions with the Gemini/NIRI adaptive optics imager \citep{hodapp2003}. Such companions can dilute the light curve, thereby biasing the measured radius, or even be the source of a false positive if the visual companion is an eclipsing binary (e.g., \citealp{ciardi2015}). We collected 9 images in the Br$\gamma$ filter, each with exposure time 5.7 s, and dithered the telescope by $\sim$2'' between each exposure. This dither pattern allows for a sky background frame to be constructed from the science images themselves. We processed the data using a custom code which performs bad pixel and flat corrections, subtracts the sky background, aligns the star between images and coadds the frames. The final image can be seen in Figure \ref{Fig:AO_NIRI}. We inspected the final image visually, and did not identify companions anywhere in the field of view, which is 26.8''$\times$26.8''. To estimate the sensitivity of these images to the presence of visual companions, we injected scaled copies of the stellar PSF at several radial separations and position angles, and scaled their brightness until each could be detected at 5-$\sigma$. Sensitivity was then averaged over position angle, and Figure \ref{Fig:AO_NIRI} shows the sensitivity to visual companions as a function of radius. We are sensitive to companions 5 mag fainter than the host beyond 270 mas, and reach a contrast of 7.3 mag in the background limited regime, beyond $\sim$1.05''.

\subsubsection{Gemini/Zorro}
\begin{figure}
   \centering
   \includegraphics[width=\hsize]{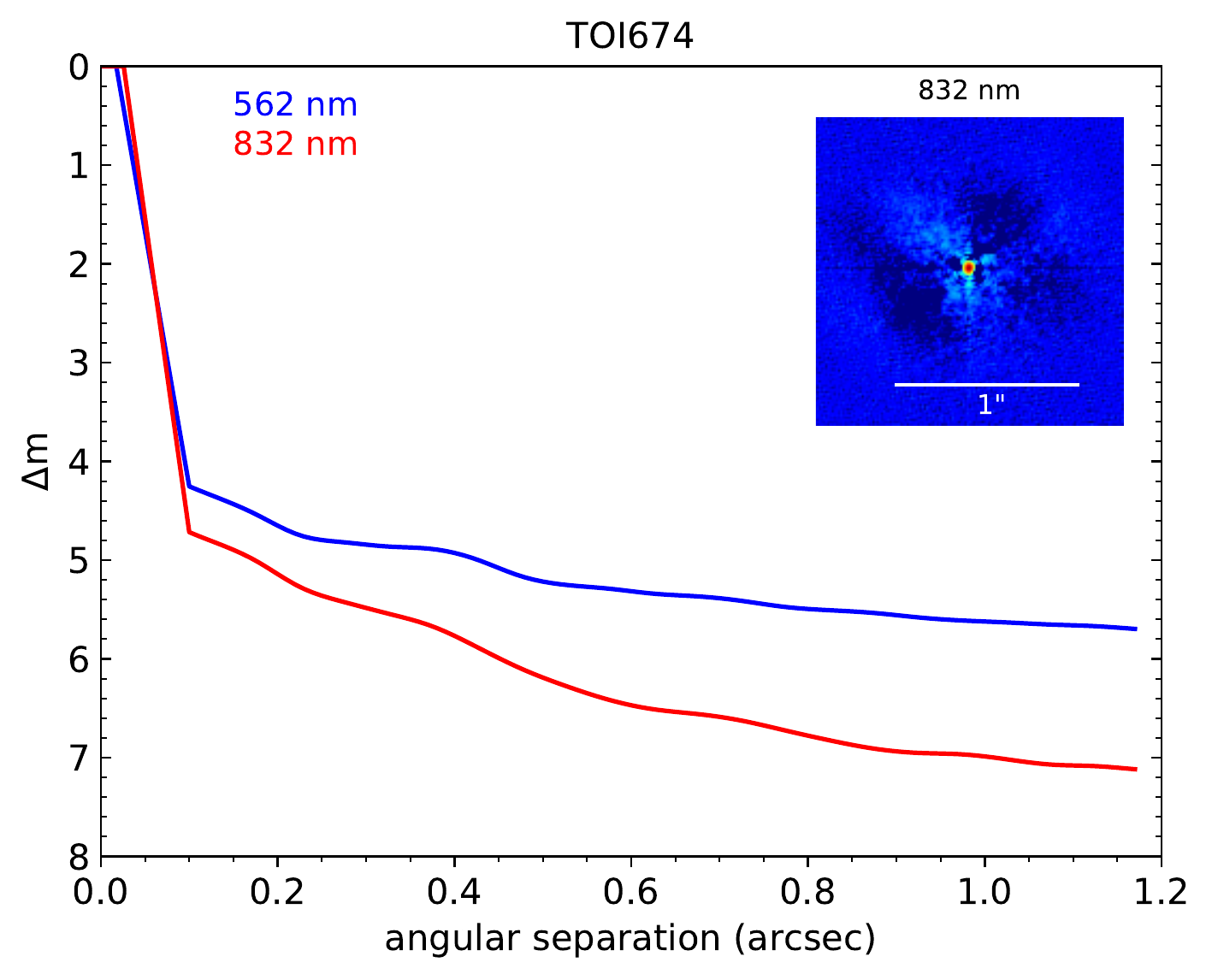}
   \caption{Gemini/Zorro high resolution image of TOI-674 taken on 14 January 2020. TOI-674 is a single star to contrast limits of 5 to 7 mag within 0.8/1.3 AU (562/832 nm respectively) out to 55 AU ($\mathrm{d}=46$ pc). \textit{Inset:} 1.2\arcsec{}x1.2\arcsec{} thumbnail image of TOI-674, centered on the star.}
   \label{Fig:AO_Zorro}
 \end{figure}

We observed TOI-674 on 14 January 2020 using the Zorro instrument mounted on the 8-m Gemini South telescope, located on Cerro Pachón in Chile. Zorro simultaneously observes diffraction-limited images at 562 nm (0.017") and 832 nm (0.028"). Our data set consisted of five 1000 X 60 ms exposures simultaneously obtained in both band-passes, followed by a single $1000 \times 60$ ms image, also in both band-passes, of a PSF standard star. Following the procedures outlined in \cite{Howell2011}, we combined all images and subjected them to Fourier analysis, and produce re-constructed imagery from which 5-$\sigma$ contrast curves are derived in each passband (Figure \ref{Fig:AO_Zorro}). Our data reveal TOI-674 to be a single star to contrast limits of 5 to 7 magnitudes within the spatial limits of 0.8/1.3 AU (562/832 nm respectively) out to 55 AU ($\mathrm{d}=46$ pc). A second Gemini/Zorro observation made in 25 February 2021 confirmed the previous result and no companion stars to TOI-674 were detected within the angular and contrast limits explored.

\section{Analysis}
\label{Sec:Analysis}

\subsection{Stellar parameters}
\label{Sec:StellarParams}
We computed an initial estimate of TOI-674 stellar parameters using the weighted average of the HARPS spectra and analysed it with \texttt{SpecMatch-Emp} (\citealp{Yee2017}). We thereby obtained $\mathrm{T_{eff}}= 3470\pm 70 $ K, $\mathrm{R_\star} = 0.413 \pm 0.100$ $\mathrm{R_\odot}$, and $[\mathrm{Fe}/\mathrm{H}] = 0.18 \pm 0.09$ dex. The parameters obtained with \texttt{SpecMatch-Emp} were used as a prior to fit the Spectral Energy Distribution (SED) of TOI-674. To estimate the stellar parameters we followed \cite{Diaz2014}. For the SED fit we adopted the distance and the apparent magnitudes from {\it Gaia} (\citealp{GaiaEDR32020}), 2MASS (\citealp{Cutri2003}), and WISE (\citealp{Cutri2014}). We used as priors the distance to the star (based on \textit{Gaia} EDR3 parallax) and the effective temperature and value of $[\mathrm{Fe}/\mathrm{H}]$ from \texttt{SpecMatch-Emp}; the uncertainty in $[\mathrm{Fe}/\mathrm{H}]$ was changed from $0.09$ dex to $0.12$ dex according to Table 3 of \cite{Yee2017}. The fit was done by comparing the SED with the synthetic spectra models from the BT-Settl library (\citealp{Allard2011}) alone and the BT-Settl library plus stellar evolution models from \cite{Dotter2008}.

The stellar parameters for TOI-674 are presented in Table \ref{Table:TOI674dinfo}. The stellar mass value for the SED fit without stellar evolution models was computed following \cite{Mann2019} that uses the distance, K-magnitude, and metallicity to obtain a mass estimate. For the SED fit using BT-Settl stellar models we find a stellar effective temperature of $3514\pm 57$ K, indicating that the spectral type must be close to M2V. To be conservative in our uncertainties, we adopted as final stellar parameters the values found by the SED fit with stellar atmosphere models without including the stellar evolution models, i.e., middle column of Table \ref{Table:TOI674dinfo}. Hence, to derive absolute planet parameters we use in this work a stellar mass and radius of $\mathrm{M}_\star=0.420\pm 0.010$ M$_\odot$ and $\mathrm{R}_\star = 0.420\pm 0.013$ R$_\odot$ respectively.

To check our results and as an independent determination of the basic stellar parameters, we performed an analysis of the broadband spectral energy distribution (SED) of the star together with the {\it Gaia\/} EDR3 parallax \citep[with no systematic correction][]{StassunTorres:2021}, in order to determine an empirical measurement of the stellar radius, following the procedures described in \citet{Stassun:2016,Stassun:2017,Stassun:2018}. We pulled the $JHK_S$ magnitudes from {\it 2MASS}, the W1--W4 magnitudes from {\it WISE}, the $G G_{\rm BP} G_{\rm RP}$ magnitudes from {\it Gaia}, and the NUV magnitude from {\it GALEX}. Together, the available photometry spans the full stellar SED over the wavelength range 0.2--22~$\mu$m (see Figure~\ref{Fig:TOI674_SED}).  

We performed a fit using NextGen stellar atmosphere models (\citealp{Hauschildt1999}), with the effective temperature ($T_{\rm eff}$), surface gravity ($\log g$), and metallicity ([Fe/H]) adopted from the spectroscopic analysis. The remaining free parameter is the extinction, $A_V$. The resulting fit (Figure~\ref{Fig:TOI674_SED}) has a reduced $\chi^2$ of 3.1 (excluding the UV measurement which appears to be in excess, see below), with best fit $A_V = 0.02 \pm 0.02$. Integrating the (unreddened) model SED gives the bolometric flux at Earth, $F_{\rm bol} = 3.50 \pm 0.12 \times 10^{-10}$ erg~s$^{-1}$~cm$^{-2}$. Taking the $F_{\rm bol}$ and $T_{\rm eff}$ together with the {\it Gaia\/} parallax, gives the stellar radius, $R_\star = 0.413 \pm 0.015$~R$_\odot$; a value consistent with our adopted stellar radius presented in Table \ref{Table:TOI674dinfo}. 

Finally, we can use the star's activity to estimate an age via empirical rotation-activity-age relations. The mean chromospheric activity from the time-series spectroscopy is $\log R'_{\rm HK} = -5.14 \pm 0.01$, which implies a stellar rotation period of $P_{\rm rot} = 52 \pm 5$~days via the empirical relations of \citet{Astudillo-Defru:2017}. This estimated $\mathrm{P}_{\rm rot}$ implies an age of $\tau_\star = 3.2 \pm 1.2$~Gyr via the empirical relations of \citet{Engle:2018}, which is consistent with the age derived from the SED fit plus stellar evolution models presented in Table \ref{Table:TOI674dinfo}.

\subsection{Frequency analysis of the HARPS data}
\begin{figure}
   \centering
   \includegraphics[width=\hsize]{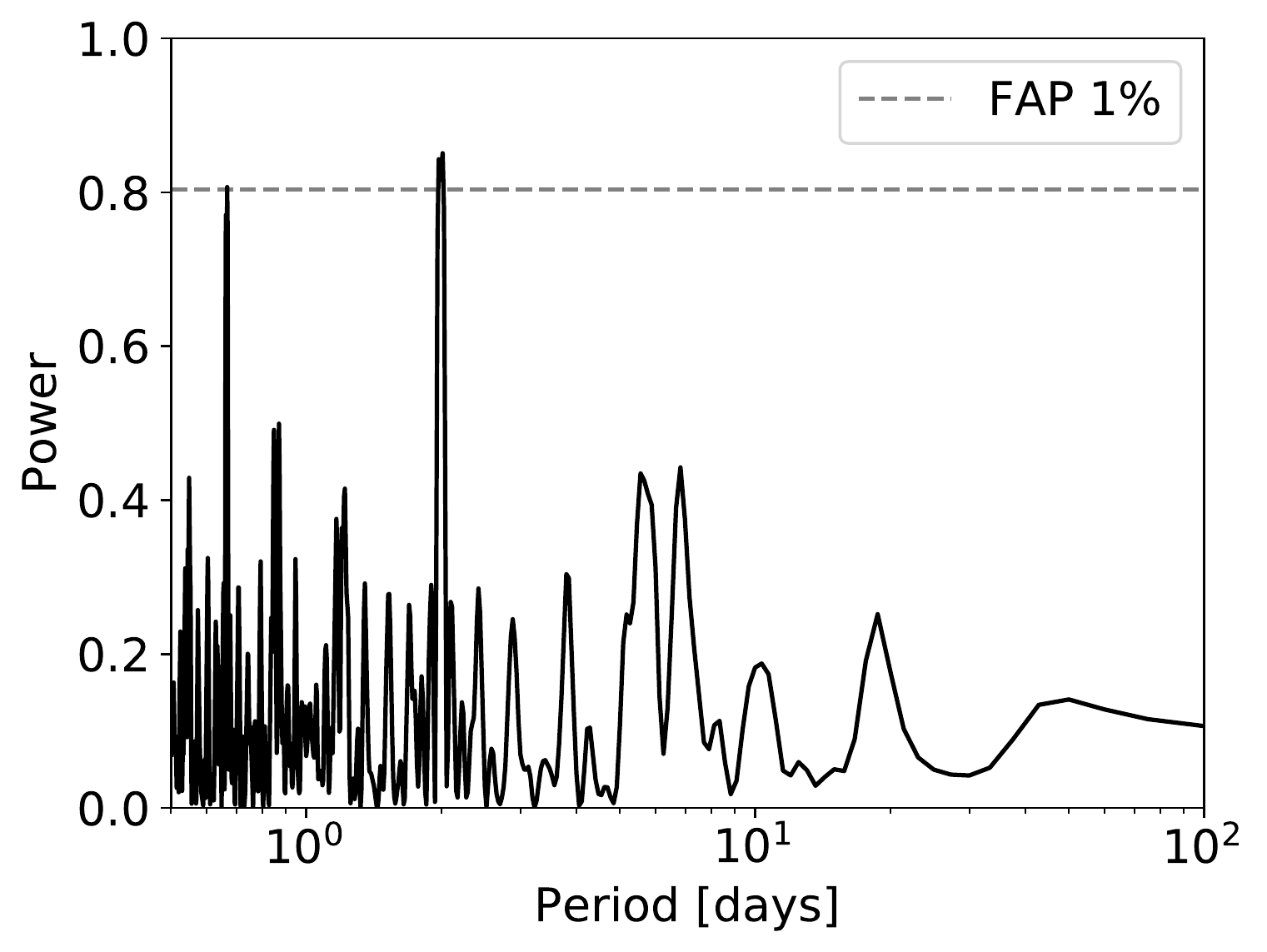}
   \caption{TOI-674b GLS periodogram for HARPS radial velocities measurements. The peak with the maximum power corresponds to a period of $\sim 2$ days, the dashed line shows the False Alarm Probability (FAP) level of 1\%.}
    \label{Fig:HARPS_Peridogram}
\end{figure}

We searched for the planetary signal in the HARPS data using a generalized Lomb-Scargle periodogram (\citealp{Zechmeister2009}). With the radial velocity measures alone a signal with a period of $\sim 2.0$ days is detected with a False Alarm Probability (FAP) below 1\% ( $ \mathrm{FAP}(peak) = 0.15\%$, see Figure \ref{Fig:HARPS_Peridogram}). This period is consistent with the orbital period originally reported by \textit{TESS} team as part of their alerts ($P = 1.977$ days).

We find no signals in the RV or activity indices periodograms that points to the rotational period of the star (see Fig. \ref{Fig:TOI674_ActivIndx_Periodogram}). M dwarfs can have rotational periods ranging from a few hours to over one hundred days (e.g., \citealp{Newton2016}) and our total baseline of HARPS observations is of $\sim 54$ days (with 17 individual spectra). In section \ref{Sec:StellarParams} we derived a rotational period for the host star of $P_{\rm rot} = 52 \pm 5$~days from empirical relations, hence it is possible that we barely observed a single full rotational period of the star during the spectroscopic follow-up of TOI-674.

\subsection{Transit light curve and radial velocity model}
The light curves were modeled with the help of the transit modeling package  \texttt{PyTransit}\footnote{\url{https://github.com/hpparvi/PyTransit}} (\citealp{Parviainen2015}) version 2.0. We used the \texttt{PyTransit} model implementation of the \cite{Mandel2002} analytic transit model with a quadratic limb darkening law.

Due to the large number of data points, the \textit{Spitzer} and ground based light curves were detrended before performing the joint fit to speed up the fitting process. For the \textit{Spitzer} time series we performed Pixel Level Decorrelation (PLD) following \cite{Deming2015} to remove the instrumental noise known to affect \textit{Spitzer} data. We computed the normalized pixel intensities for each of the $3\times 3$ \textit{Spitzer} pixels 

\begin{equation}
    \hat{P}_i^t = \frac{P_i^t}{\sum_{i=1}^N P_i^t}
\label{Eq:PixIntensities}
\end{equation}
where $P_i^t$ is the $i$-th pixel inside the aperture at a time $t$ of the time series. We modeled the total flux inside an aperture as 

\begin{equation}
    S^t = \sum_{i=1}^N c_i \hat{P}_i^t + T(t) + ft + gt^2
\label{Eq:PLD}
\end{equation}
where $c_i$ are weighting coefficients for the normalized pixel intensities $\hat{P}_i^t$, $T(t)$ is the transit model, and $ft + gt^2$ is a quadratic function in time with constants $f$ and $g$. We fitted the \textit{Spitzer} time series using equation \ref{Eq:PLD} and compute the $\chi^2$ for different aperture sizes and binning the data in time with several bin sizes. The combination of aperture and bin sizes that delivered the lowest $\chi^2$ value was used for the global fit.

The ground based light curves were fitted simultaneously using a common transit model and the systematic effects of each data set were accounted for using a linear model with two free parameters dependent on the star position in the detector (X and Y-axis), a term dependent on the Full Width at Half Maximum (FWHM) of the PSF as a proxy for seeing variations, and a time-dependent term to model any time-dependent slope present in the time series.

For the joint fit, the global free parameters for the transit modeling included the planet-to-star radius ratio $\mathrm{R}_\mathrm{p}/\mathrm{R}_\star$, the central time of the transit $\mathrm{T}_{\mathrm{c}}$, the stellar density $\rho_\star$, and the transit impact parameter $\mathrm{b}$. The quadratic limb darkening (LD) coefficients $\mathrm{u}_1$ and $\mathrm{u}_2$ were set free, but during the fit these values were compared to the predicted coefficients computed for each bandpass using \texttt{ldtk}\footnote{\url{https://github.com/hpparvi/ldtk}} (\citealp{Parviainen2015b}). \texttt{Ldtk} uses the \cite{Husser2013} spectral library to compute custom stellar limb darkening profiles. For each light curve the predictd quadratic LD coefficients were computed using the stellar parameters derived in section 3.1 and the fitted coefficients were weighted against these predicted values using a likelihood function. During the fitting process we converted the LD coefficients $(\mathrm{u}_1, \mathrm{u}_2)$ to the parameterization proposed by \cite{Kipping2013} $(\mathrm{q}_1,\mathrm{q}_2)$ with $q_i \in [0,1]$.

For the \textit{TESS} time series, besides using an analytical model for the transit we fitted the stellar variability present in the time series using Gaussian Processes (GPs; e.g., \citealp{Rasmussen2006}, \citealp{Gibson2012}, \citealp{Ambikasaran2015}). The \textit{TESS} GPs were computed using the \texttt{python} package \texttt{Celerite} (\citealp{ForemanMackey2017}); we chose a Matern $3/2$ kernel:

\begin{equation}
    k_{ij\; \mathrm{TESS}} = c^2_1 \left( 1 + \frac{\sqrt{3} |t_i-t_j|}{\tau_1}\right) \exp\left(-\frac{\sqrt{3} |t_i-t_j|}{\tau_1}\right)
\label{Eq:TESS_GPKernel}
\end{equation}
where $|t_i-t_j|$ is the time between points in the series, $c_1$ is the amplitude of the variability, and $\tau_1$ is a characteristic time-scale. The constants $c_1$ and $\tau_1$ were set as free parameters.

The radial velocity data was modeled using \texttt{RadVel}\footnote{\url{https://github.com/California-Planet-Search/radvel}} (\citealp{Fulton2018}). The free parameters in the radial velocity fit were the planet induced radial velocity semi-amplitude ($K_\mathrm{RV}$), the host star systemic velocity ($\gamma_0$), the instrumental radial velocity jitter ($\sigma_{\mathrm{RV\; jitter}}$), while the orbital period and central transit time were also set free but taken to be global parameters in common with the fits of the light curves. To account for systematic noise present in the radial velocity time series we use GPs with an exponential squared kernel (i.e., a Gaussian kernel)

\begin{equation}
    k_{ij\; \mathrm{RV}} = c^2_2 \exp \left(  - \frac{ (t_i-t_j)^2}{\tau^2_2} \right)
\label{Eq:RV_GPKernel}
\end{equation}
where $t_i-t_j$ is the time between points in the series, $c_2$ is the amplitude of the exponential squared kernel, and $\tau_2$ is a characteristic time-scale. The constants $c_2$ and $\tau_2$ were set as free parameters.

In order to estimate the fitted parameter values we employed a Bayesian approach. We started the fitting procedure by doing an uninformative transit search in the \textit{TESS} time series (from Sector 9 and 10) using Transit Least Squares (TLS, \citealp{Hippke2019}). From TLS we obtained an estimate of the period and epoch of the central time of the transit with their respective uncertainties, we used these values as priors for the joint fit. Then, we implemented a Markov chain Monte Carlo (MCMC) procedure using \texttt{emcee} (\citealp{ForemanMackey2013}) to evaluate a likelihood plus a prior function. The likelihood function was the sum of the log likelihood for each transit time series and the log likelihood of the radial velocity observations. A total of 38 free parameters were sampled in the joint fit.

The fitting procedure started with a global maximization of the posterior function using \texttt{PyDE}\footnote{\url{https://github.com/hpparvi/PyDE}}. Once the minimization converged we launched a burn-in MCMC with 125 chains and 2000 iterations. After this burn-in stage was finished, the main MCMC ran for 5000 iterations with the same number of chains as the burn-in stage. The final parameter values and 1$-\sigma$ uncertainties were determined from the posterior distributions of the fitted variables: we computed the percentiles of the distribution corresponding to the median and lower and upper 1-$\sigma$ limits (from the median) of the distribution for each variable.

A planet in such a short orbital period ($P \sim 1.9$ days) is likely to have had its orbit circularized over time. However, this is not necessarily true. There is evidence that Neptune-mass planets with orbital periods of the order of a few days present non-zero orbital eccentricity (e.g., \citealp{Kane2012}, \citealp{Correia2020}). Thus, we performed two global fits: one assuming a circular orbit and the other allowed the orbit to have non-zero eccentricity. For the case of the eccentric orbit we set as global free parameters the square root of the eccentricity multiplied by the sine of the argument of the periastron (i.e., $\sqrt{e}\sin(\omega)$ with parameter limits $[-1,1]$) and the square root of the eccentricity multiplied by the cosine of the argument of the periastron (i.e., $\sqrt{e}\cos(\omega)$ with parameter limits $[-1,1]$). Using this parametrization we are sampling values of $e \in [0,1]$ and $\omega \in [0, 2\pi]$ (after imposing $e < 1$).

Table \ref{Table:TOI674b_ParamsPriors} presents the prior functions and limits used in both global fits. For the circular orbit case we used an uninformative prior (uniform) for the stellar density, with $\rho_\star \in [5,15]$ g cm$^{-3}$. For the fit allowing a non-zero eccentricity, we imposed a more restrictive range of values for the stellar density, using a normal prior with the density and 2-$\sigma$ uncertainty found by our procedure used to obtain the stellar parameters (see Section \ref{Sec:StellarParams}).   


\section{Results and discussion}
\label{Sec:Results}

\begin{figure*}
   \centering
   \includegraphics[width=\textwidth]{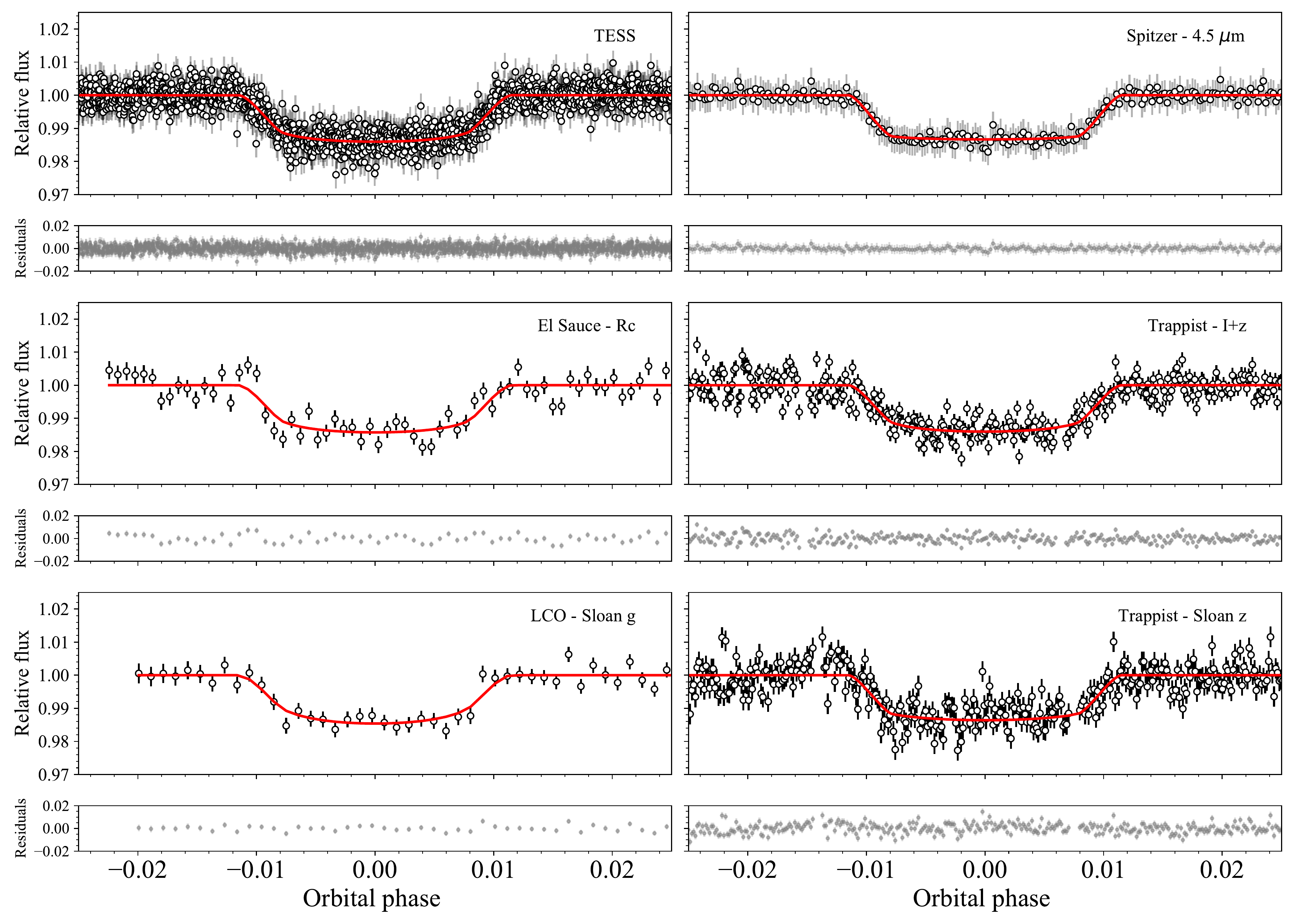}
   \caption{TOI-674b space and ground-based transit light curves after removing systematic noise affecting the data. The best circular orbit model fit is represented by the red line and below each panel the residuals of the fit are shown.}
    \label{Fig:LightCurves}%
\end{figure*}

\begin{figure*}
   \centering
   \includegraphics[width=\textwidth]{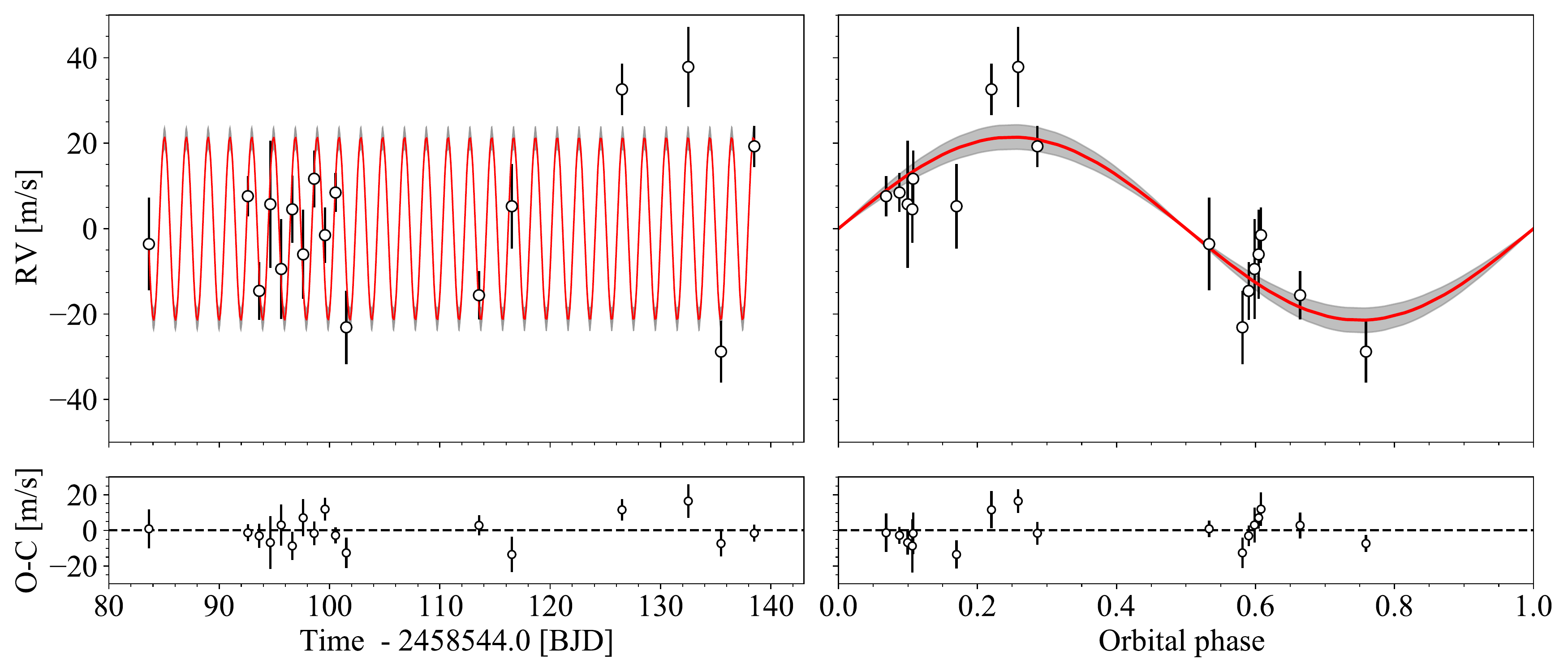}
   \caption{TOI-674b HARPS radial velocity measurements and best fitting circular orbit model including Gaussian processes to model systematic noise sources (red line). \textit{Left}: HARPS RV time series measurements. \textit{Right}: HARPS RV measurements phase-folded to the orbital period of the planet.The shaded gray area corresponds to $\pm 1$-$\sigma$ level from the mean of 5000 randomly selected posterior samples. The RMS of the residuals is $8.25$ m/s.}
    \label{Fig:HARPS_RV}
\end{figure*}

The global fit parameter values and 1-$\sigma$ uncertainties are presented in Table \ref{Table:TOI674b_Params}. Figures \ref{Fig:Fit_Circular_CornerPlot} and \ref{Fig:Fit_Eccentric_CornerPlot} present the correlation plots of the orbital fitted parameters excluding the limb darkening coefficients and parameters related to the red noise. Figures \ref{Fig:LightCurves} and \ref{Fig:HARPS_RV} show all the light curves included in this study and the radial velocity measurements made by HARPS; the best circular orbit model fit from our joint modeling is shown in red.

\begin{table*}
  \caption[]{TOI-674b global fit parameters and 1-$\sigma$ uncertainties for the circular and non-zero eccentricity fits. The prior functions and parameter limits used in the joint fit are presented in Table \ref{Table:TOI674b_ParamsPriors}. The adopted stellar mass and radius used to derive the absolute planet parameters are the ones presented in the middle column of Table \ref{Table:TOI674dinfo} (i.e., BT-Settl models SED fit).}
  \label{Table:TOI674b_Params}
  \centering
  \begin{tabular}{lcc}
    \hline \hline
    Fitted orbital parameter &  Circular orbit ($e=0$) & Non circular orbit ($e \neq 0)$ \\
    \hline

$R_{p}/R_\star$ & $0.1140 \pm 0.0009$ & $0.1143 \pm 0.0009$ \\
$T_{\mathrm{c\;BJD}}$ [days] & $2458641.404552 \pm 0.000102$ & $2458641.404604 \pm 0.000108$ \\
$P$ [days] & $1.977143 \pm 0.000003$ & $1.977143 \pm 0.000003$ \\
$\rho_\star$ [g cm$^{-3}$] & $10.14 \pm 1.04$ & $7.97 \pm 0.59$ \\
$b = (a/R_\star) \cos(i) \left( \frac{1-e^2}{1+e\sin(\omega)} \right)$ & $0.624 \pm 0.035$ & $0.635 \pm 0.030$ \\
$\sqrt{e}\cos(\omega)$ & ----  & $0.217 \pm 0.153$ \\
$\sqrt{e}\sin(\omega)$ & ----  & $0.211 \pm 0.089$ \\
$\gamma_0$ [m/s] & $13485.51 \pm 4.78$ & $13486.59 \pm 3.81$ \\
$K$ [m/s] & $21.44 \pm 2.94$ & $23.55 \pm 3.70$ \\
$\sigma_{\mathrm{RV}}$ [m/s] & $3.40 \pm 2.95$ & $3.58 \pm 2.85$ \\
\hline
Derived orbital parameters & & \\
\hline
$e$ & ----  & $0.10 \pm 0.05$ \\
$\omega$ [deg] & ----  & $45.4 \pm 30.8$ \\
$a/R_\star$ & $12.80 \pm 0.42$ & $11.81 \pm 0.30$ \\
$i$ [deg] & $87.21 \pm 0.24$ & $86.68 \pm 0.20$ \\
\hline
Fitted LD coefficients & & \\
\hline
$\mathrm{u}_{\mathrm{1\;TESS}}$ & $0.25 \pm 0.13$ & $0.24 \pm 0.13$ \\
$\mathrm{u}_{\mathrm{2\;TESS}}$ & $0.36 \pm 0.19$ & $0.36 \pm 0.17$ \\
$\mathrm{u}_{\mathrm{1\;4.5\mu m}}$ & $0.071 \pm 0.036$ & $0.073 \pm 0.043$ \\
$\mathrm{u}_{\mathrm{2\;4.5\mu m}}$ & $0.15 \pm 0.07$ & $0.15 \pm 0.07$ \\
$\mathrm{u}_{\mathrm{1\;R_c}}$ & $0.29 \pm 0.18$ & $0.30 \pm 0.17$ \\
$\mathrm{u}_{\mathrm{2\;R_c}}$ & $0.42 \pm 0.26$ & $0.41 \pm 0.24$ \\
$\mathrm{u}_{\mathrm{1\;I+z}}$ & $0.24 \pm 0.13$ & $0.23 \pm 0.14$ \\
$\mathrm{u}_{\mathrm{2\;I+z}}$ & $0.36 \pm 0.19$ & $0.36 \pm 0.19$ \\
$\mathrm{u}_{\mathrm{1\;g}}$ & $0.48 \pm 0.26$ & $0.48 \pm 0.24$ \\
$\mathrm{u}_{\mathrm{2\;g}}$ & $0.38 \pm 0.30$ & $0.38 \pm 0.29$ \\
$\mathrm{u}_{\mathrm{1\;z}}$ & $0.38 \pm 0.13$ & $0.39 \pm 0.15$ \\
$\mathrm{u}_{\mathrm{2\;z}}$ & $-0.01 \pm 0.19$ & $-0.04 \pm 0.21$ \\
\hline
\textit{TESS} GP parameters & & \\
\hline
$\mathrm{c}_1$ & $(2.1 \pm 0.4) \times 10^{-4}$ & $(2.1 \pm 0.4) \times 10^{-4}$ \\
$\tau_1$ [days] & $0.74 \pm 0.30$ & $0.75 \pm 0.35$ \\
\hline
HARPS GP parameters & & \\
\hline
$\mathrm{c}_2$ [m/s] & $4.3^{+9.9}_{-3.2}$ & $3.3^{+5.5}_{-2.4}$ \\
$\tau_2$ [days] & $49 \pm 31$ & $45 \pm 30$ \\
\hline
Derived planet parameters & & \\
\hline
$\mathrm{R}_\mathrm{p}$ [$\mathrm{R}_\oplus$] & $5.25 \pm 0.17$ & $5.26 \pm 0.17$ \\
$\mathrm{M}_\mathrm{p}$ [$\mathrm{M}_\oplus$] & $23.6 \pm 3.3$ & $23.5 \pm 3.3$ \\
$\rho_\mathrm{p}$ [g cm$^{-3}$] & $0.91 \pm 0.15$ & $0.90 \pm 0.15$ \\
$\mathrm{g}_\mathrm{p}$ [m s$^{-2}$] & $8.5 \pm 1.3$ & $8.4 \pm 1.3$ \\
$\mathrm{a}$ [AU] & $0.0250 \pm 0.0008$ & $0.0231 \pm 0.0007$ \\
$\mathrm{T}_\mathrm{eq}$ [K]\tablefootmark{1} & $635 \pm 15$ & $661 \pm 14$ \\
$\langle \mathrm{F}_\mathrm{p} \rangle$ [10$^5$ W/m$^2$] & $0.523 \pm 0.039$ & $0.612 \pm 0.045$ \\
$\mathrm{S}_\mathrm{p}$ [$\mathrm{S}_\odot$] & $38.73 \pm 2.88$ & $45.48 \pm 3.39$ \\
\hline
\end{tabular}
\tablefoot{$\mathrm{T}_\mathrm{eq}$ computed assuming an albedo of 0.3.}
\end{table*}

\subsection{The orbital eccentricity of TOI-674b}
By using a normal prior on the stellar density with the values taken from the stellar parameters for the host star and setting the eccentricity and argument of the periastron free, we find that TOI-674b may have a slightly eccentric orbit with $\mathrm{e} = 0.10 \pm 0.05$, although this eccentricity value could be an spurious signal. According to \cite{Lucy1971}, an eccentricity should not be considered significant if it is less than 2.45$\sigma$ from zero, which our eccentric fit does not achieve. Nonetheless, many Neptune-mass planets with short orbital periods posses non-circular orbits like HAT-P-11b (\citealp{Bakos2010}, \citealp{Yee2018}), GJ 436b (\citealp{Gillon2007}, \citealp{Lanotte2014}), and TOI-1728b (\citealp{Kanodia2020}) to name some examples. \cite{Correia2020} proposed that these non-circular orbits could be produced by any of several processes alone or in combination like a tidal torque created by photo-driven evaporation of the planet's atmosphere and/or gravitational interaction between the short period planet and another planetary object in a longer period orbit in the system.

We compared the Bayesian Information Criterion (BIC) statistics of the circular and eccentric orbit model fits to the data. The BIC takes into account the number of observed data points $n$, the number of fitted parameters $k$, and the maximized likelihood $\mathcal{L}_\mathrm{max}$. The BIC is defined by

\begin{equation}
    \mathrm{BIC} = k \ln{n} - 2\ln{\mathcal{L}_\mathrm{max}}\;.
\label{Eq:BIC}
\end{equation}
The model with the lowest BIC value is the one that better fits the data. In our case $\Delta \mathrm{BIC} = \mathrm{BIC}_\mathrm{Ecc}- \mathrm{BIC}_\mathrm{Circ} = 44$, meaning that the circular model fit is preferred using this criterion. 

It is worth pointing out that the stellar density $\rho_\star$, transit impact parameter $\mathrm{b}$, eccentricity $\mathrm{e}$, and argument of the periastron $\omega$ are correlated. In our non-zero eccentricity model fit we used a normal prior to the stellar density which could have skewed our results to eccentricity values higher than zero. Another point to consider is that our radial velocity measurements were made with a rather nonuniform distribution in orbital phase, in part due to the orbital period being almost an integer number of days ($P \sim 1.9$ days) making a uniform coverage difficult from a single observing site. The non-uniform phase coverage contributes to the difficulty of establishing whether the orbit is slightly eccentric. Additional RV measurements in different orbital phases and/or observations with higher precision, for example using VLT/ESPRESSO (\citealp{Pepe2010,Pepe2021}), could help to put stronger constraints on the orbital eccentricity of TOI-674b.

On the other hand, the stellar density found by our circular orbit fit using a uniform prior for this parameter is $\rho_\star = 10.14 \pm 1.04$ g cm$^{-3}$, a value that differs from the derived stellar density computed using our adopted stellar mass and radius ($\rho_\star = 7.99 \pm 0.76$ g cm$^{-3}$) by $\sim 2 \sigma$. This discrepancy could be caused by an underestimation of the uncertainties of the stellar parameters or an unknown systematic affecting our fit. The different stellar density values found in the circular and non-circular orbit fit affected the values of the semi-major axis over stellar radius ($a/R_\star$) and orbital inclination ($i$) (see Table \ref{Table:TOI674b_Params}) since these values were derived using the stellar density, impact parameter, eccentricity and argument of the periastron.

Despite the different values of stellar density from our circular and non-zero eccentricity results, both cases are in agreement within the 1-$\sigma$ uncertainties (see Table \ref{Table:TOI674b_Params}). Given the BIC results, for the rest of this work we will adopt the planetary parameters derived from the circular orbit model fit.

\subsection{Transit timing variations}
\begin{figure}
   \centering
   \includegraphics[width=\hsize]{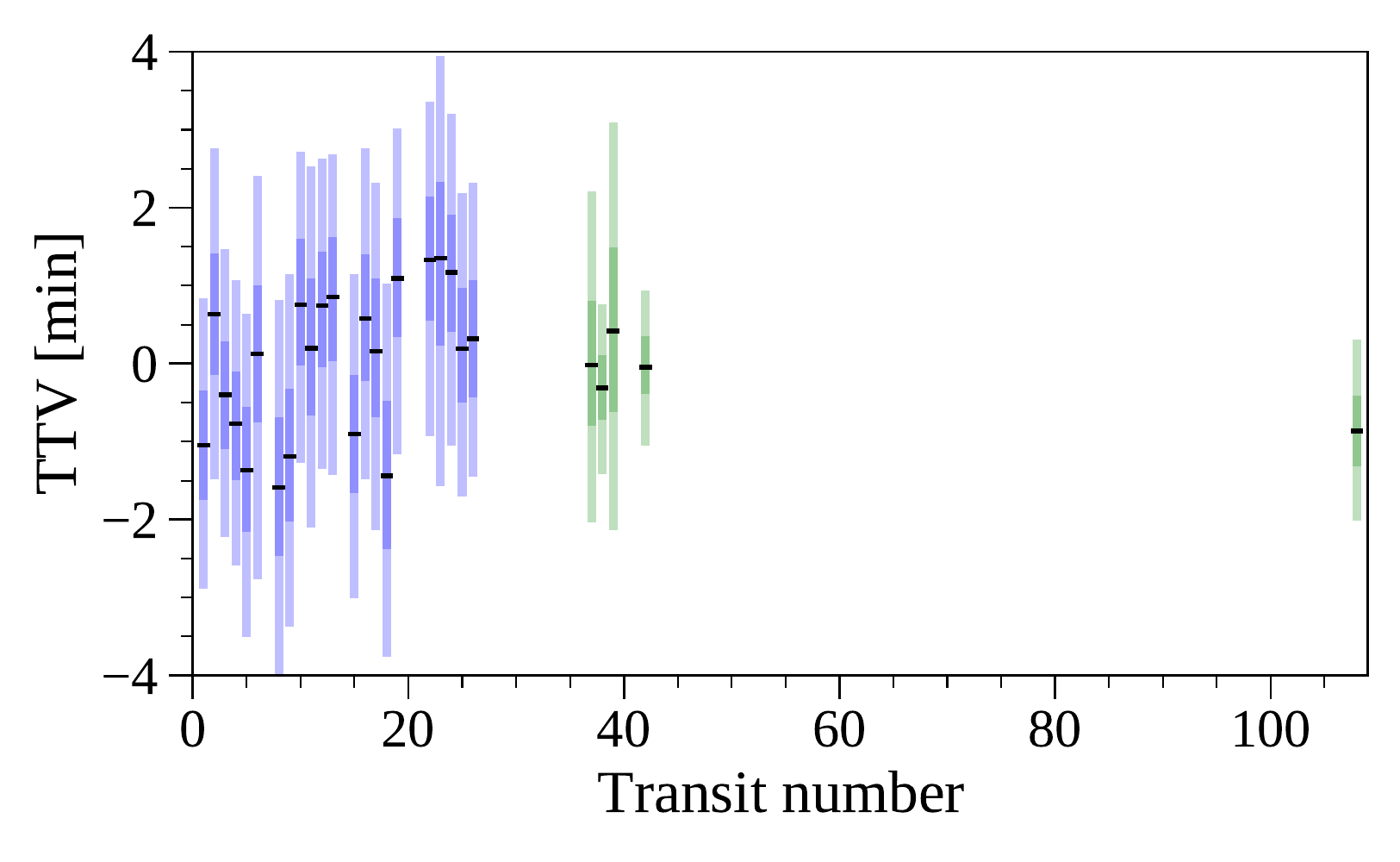}
   \caption{TOI-674b transit timing variations from \textit{TESS} data (blue) and follow-up observations (green). The different shades of color of the error bars represent the 1-$\sigma$ (dark) and 3-$\sigma$ (light) levels of uncertainty of the measurements. No significant TTVs are detected in a $\sim 200$ days of baseline covered by the observations.}
    \label{Fig:TOI674b_TTV}
\end{figure}

The presence of additional planets in the system could perturb the orbit of TOI-674b and produce an orbit with non-zero eccentricity. The interactions between TOI-674b and these hypothetical additional planets in the system could also affect the central times of the observed transits. We searched for transit timing variations (TTVs) of TOI-674b by fitting the \textit{TESS} photometry from Sector 9 and Sector 10, the \textit{Spitzer} transit, and the 4 ground-based transits jointly with \texttt{PyTTV} (\citealp{Korth2020}). This translated into 27 individual transit fits spanning a time baseline of close to $\sim 200$ days. The fitted model included individual transit centers as free parameters and shared the remaining geometric and orbital parameters such as impact parameter, radius ratio, and stellar density. We estimated the model parameter posteriors using MCMC-sampling (\texttt{emcee}; \citealp{ForemanMackey2013}). The transit centers do not show significant variations from the linear ephemeris. We show the TTVs in Fig. \ref{Fig:TOI674b_TTV}.

The photometric measurements do not cover a long enough period ($\sim 200$ days) to allow us to put further constraints on hypothetical additional planets in the system which could be responsible for an eccentric orbit of TOI-674b via gravitational interactions. Further monitoring (spectroscopic and photometric) is therefore desirable. 

\subsection{Mass, radius and composition}
We find that TOI-674b has a radius of $5.25 \pm 0.17$ $\mathrm{R}_\oplus$ and a mass of $23.6 \pm 3.3$ $\mathrm{M}_\oplus$. Combining these measurements we find that TOI-674b has a mean density of $\rho_\mathrm{p} = 0.91 \pm 0.15$ g cm$^{-3}$.  

\begin{figure*}
   \centering
   \includegraphics[width=17cm]{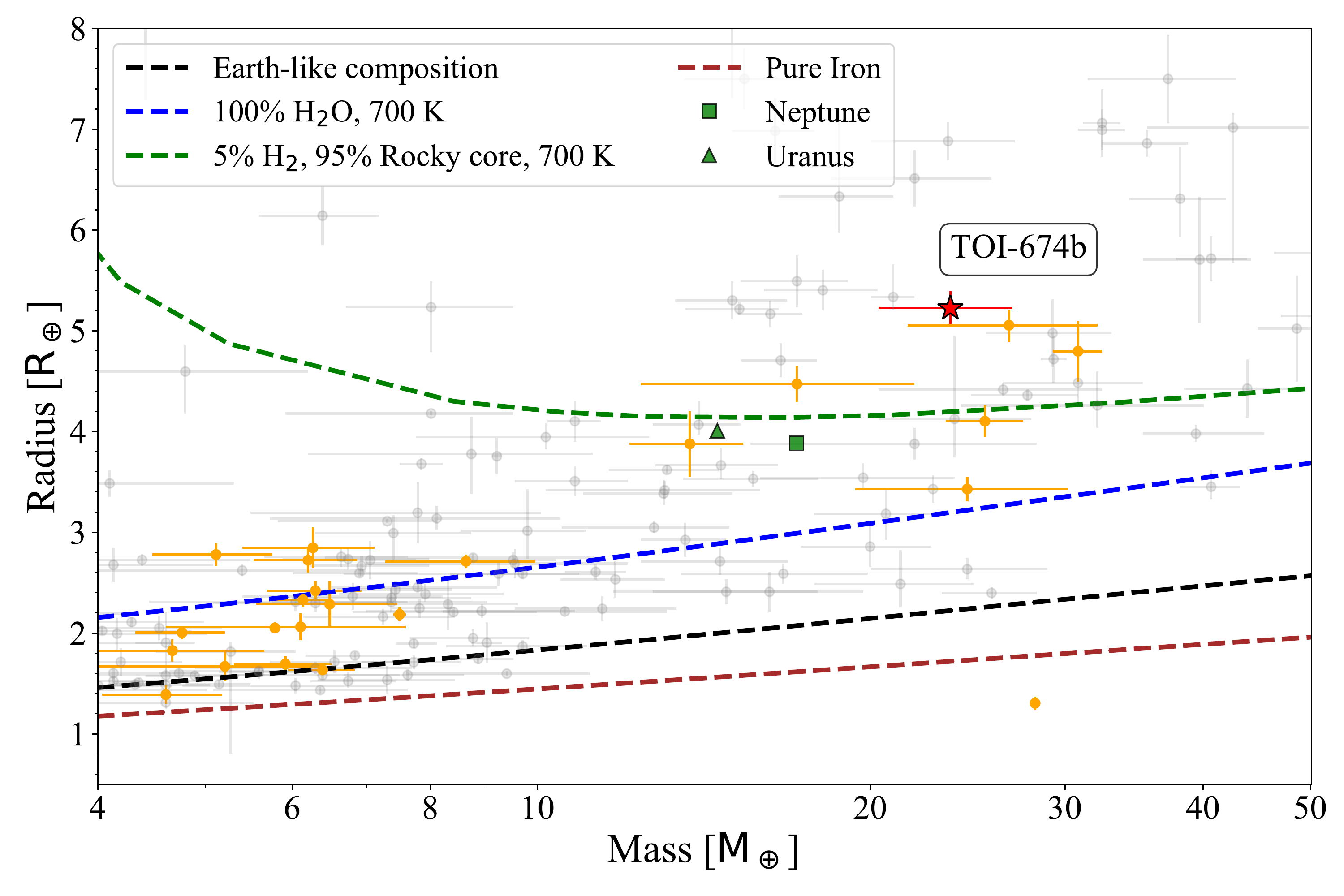}
   \caption{Mass-radius diagram for planets with mass determinations with a precision better than 30\%. The red star is TOI-674b, the orange points are planets orbiting M type stars, and the gray points are planets around other types of stars. The dashed lines represent the mass-radius models of \cite{Zeng2019}: planets with pure iron cores (100\% Fe, brown line), earth-like rocky compositions (32.5\% Fe plus 67.5\% MgSiO$_3$, black line), 100\% water compositions (blue line), and earth-like composition planet cores with 5\% H$_2$ gaseous envelopes (green line).}
    \label{Fig:TOI674b_MassRadius}
\end{figure*}

We compared the mass and radius of TOI-674b with known exoplanets taken from the TEPcat database (\citealp{Southworth2011}) and the composition models of \cite{Zeng2019}. The planet models consider an isothermal atmosphere and are truncated at 1 milli-bar pressure level (defining the radius of the planet). Figure \ref{Fig:TOI674b_MassRadius} shows the mass-radius diagram for known transiting exoplanets with masses measured with a precision better than 30\% and the rocky planet models of \cite{Zeng2019} with an equilibrium temperature of 700 K,  i.e., the equilibrium temperature closest to the likely temperature of TOI-674b. The planet models shown in Fig. \ref{Fig:TOI674b_MassRadius} are planets with pure iron cores (100\% Fe), earth-like rocky compositions (32.5\% Fe plus 67.5\% MgSiO$_3$), a 100\% water compositions, and earth-like composition planet cores with 5\% H$_2$ gaseous envelopes. We can see that TOI-674b is located far above all of these model predictions, suggesting that this planet must possess a very large H/He envelope.

Comparing the mass and radius of TOI-674b with other planets discovered orbiting around M type stars, we find that TOI-674b is one of the largest and most massive super-Neptune class planets discovered around an M-dwarf to date. The planets orbiting M-dwarfs closest to TOI-674b in the M-R diagram are TOI-1728b (\citealp{Kanodia2020}) and TOI-442b (\citealp{Dreizler2020}). TOI-1728b orbits an M0V star ($\mathrm{T}_{\mathrm{eff}} = 3980 $ K) and is a super-Neptune with $\mathrm{R}_\mathrm{p} = 5.05 \pm 0.17$ $\mathrm{R}_\oplus$, $\mathrm{M}_\mathrm{p} = 26.78 \pm 5.43$ $\mathrm{M}_\oplus$ and an orbital period of $\sim 3.5$ days ($\mathrm{T}_{\mathrm{eq}} = 767 $ K). TOI-442b is another Neptune-like planet orbiting an M0V star with $\mathrm{R}_\mathrm{p} = 4.7 \pm 0.3$ $\mathrm{R}_\oplus$, $\mathrm{M}_\mathrm{p} = 30.8 \pm 1.5$ $\mathrm{M}_\oplus$, and an orbital period of $\sim 4$ days. TOI-674b, TOI-1728b and TOI-442b have very similar masses and radii; neither planet has a very eccentric orbit. All host stars are main-sequence M dwarfs, but the star TOI-674 is smaller and less luminous than TOI-1728 and TOI-442, therefore even though its planet has a closer-in orbit, it has a slightly smaller equilibrium temperature (for the same albedo). The similarity between these three recent \textit{TESS} discoveries means that most of the qualitative information in the Discussion and Summary sections of \cite{Kanodia2020} and \cite{Dreizler2020} also applies to TOI-674b.

The voluminous envelope of TOI-674b suggests that it has more H$_2$ than could be obtained from accretion and degassing of solid material (\citealp{Rogers2011}); therefore it must have accreted gas directly from its protoplanetary disk. Accretion of large gaseous envelopes so close to a star requires a very massive solid core, so this planet very likely formed much farther from its star and then moved inwards, either via gradual disk migration early in the system's history when the protoplanetary disk was still massive or later via high-eccentricity migration and tidal circularization. Early arrival at its current orbit would have subjected TOI-674b to a long epoch of substantial photoionizing radiation, stripping the outer part of the atmosphere; tidal heating would also have induced substantial mass loss. Thus, TOI-674b probably formed with a substantially more massive envelope that was partially stripped away, mostly in the first tens of millions of years after reaching its current orbit (e.g., check \citealp{Fulton2017}, \citealp{Fulton&Petigura2018} and references therein). 

\subsection{TOI-674b and the Neptunian desert}
\begin{figure}
   \centering
   \includegraphics[width=\hsize]{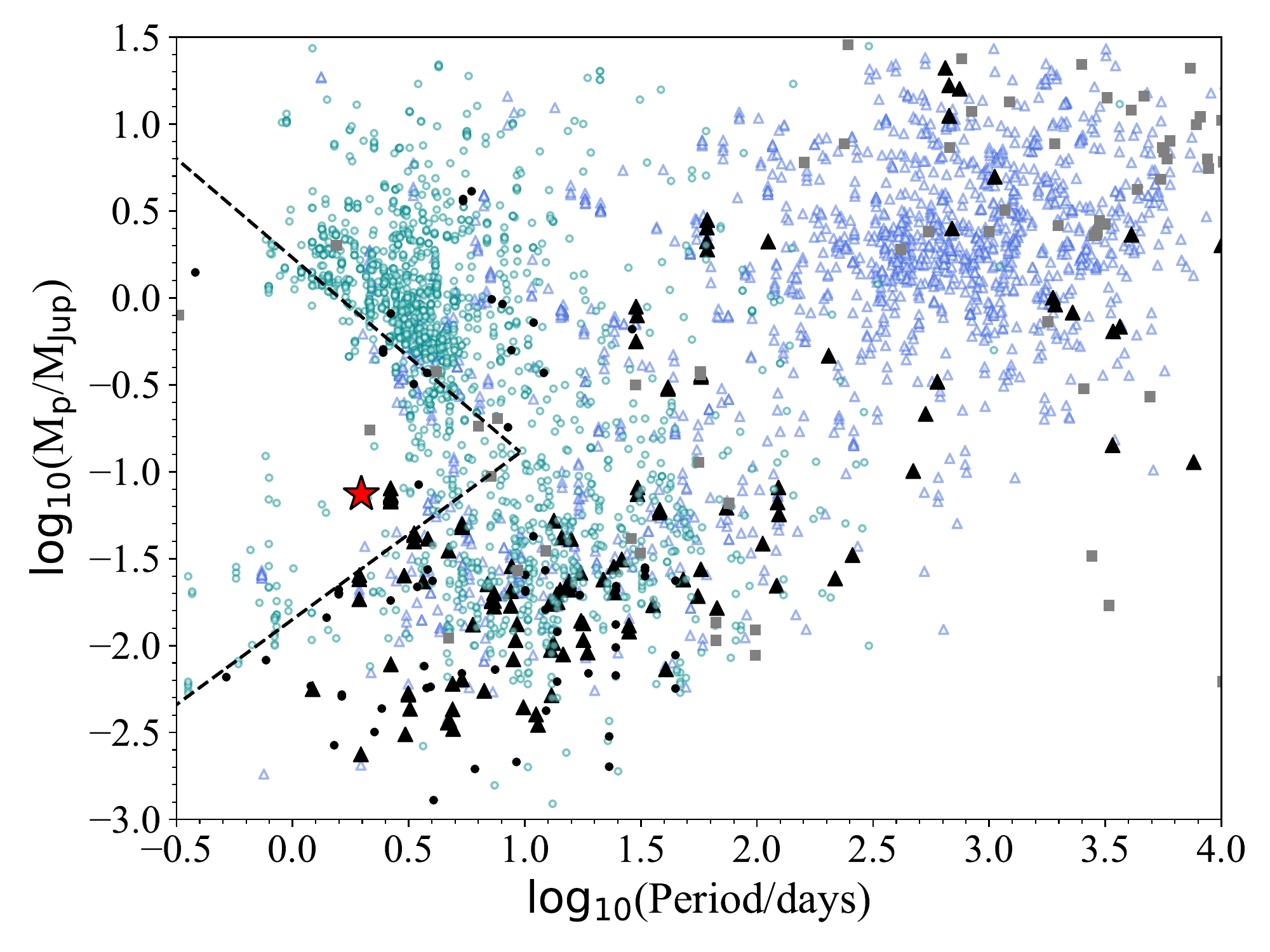}
   \caption{Mass versus orbital period diagram for planets with mass measurements. Planets found by the transit method are shown in light cyan circles, planets found by radial velocity measurements are shown in blue triangles, and planets found by other methods are shown in gray squares. The filled black symbols represent planets orbiting around M stars (i.e, stars with $\mathrm{T}_\mathrm{eff}$ in the range 2500-4000 K), each symbol representing the discovery method as explained before. The black dashed lines represent the limits of the Neptunian Desert defined by \cite{Mazeh2016}. TOI-674b position in this diagram is marked by the red star.}
    \label{Fig:TOI674b_MassPeriod}
\end{figure}

Based on our derived mass and orbital period, TOI-674b is a new resident of the so-called Neptunian Desert. In Figure \ref{Fig:TOI674b_MassPeriod} we plotted the $\log \mathrm{P}$ vs $\log \mathrm{M_p}$ relationship for known exoplanets; the black lines are the limits of the Neptunian Desert by \cite{Mazeh2016}. TOI-674b is well within the limits of the desert and in a region sparsely populated in the orbital period versus radius space.

There are multiple scenarios that explain the origin of the Neptunian desert. One proposed explanation of the lack of Neptune-sized planets in short orbits is photo-evaporation of the planet's H/He envelope due to high energy radiation coming from the star once the planet arrives to a position close to its central star after formation (e.g., \citealp{Owen2013}, \citealp{Lopez2013}, \citealp{Chen2016}). The lower boundary of the desert in \cite{Mazeh2016} (i.e., planets with low mass/small radius) is consistent with this proposed photo-evaporation mechanism. \cite{Matsakos2016} proposed that tidal disruption of planets in highly-eccentric orbits (followed by circularisation of the orbit due to tidal interactions with the host star) can explain both boundaries of the desert. \cite{Owen2018} proposed that both mechanisms are able to explain the Neptunian desert, with photo-evaporation creating the low mass/small radius and high mass/large radius boundary being created by the tidal disruption limit for gas giants experiencing high-eccentric migration. Although TOI-674b is inside the proposed boundary by \cite{Mazeh2016}, it is closer to the low mass/small radius limit indicating that this planet may have experienced some photo-evaporation of its upper atmospheric layers due to the high energy radiation of its host star.

It is also worth noting that there are few large planets known to orbit M stars. \cite{Bonfils2013} estimated the occurrence of planets around M dwarfs using HARPS data and showed that planets with large masses were not common around these types of stars. \cite{Bonfils2013} found that the occurrence for planets with $\mathrm{M}_\mathrm{p}\sin(i)$ in the range of [10, 100] $\mathrm{M}_\oplus$ and with orbital periods shorter than 10 days is $0.03^{+0.04}_{-0.01}$. A similar result was obtained by \cite{Dressing2015} using Kepler data to estimate the rate of planets around M stars. For the largest planet size range studied by \cite{Dressing2015}, i.e., $\mathrm{R}_\mathrm{p} \in [3.5,4.0]$ $\mathrm{R}_\oplus$ and orbital periods shorter than 50 days, \cite{Dressing2015} find an occurrence rate of $0.016^{+0.018}_{-0.070}$ planets per M dwarf. 

\subsection{Potential for atmospheric characterisation}
\begin{figure*}
   \centering
   \includegraphics[width=17cm]{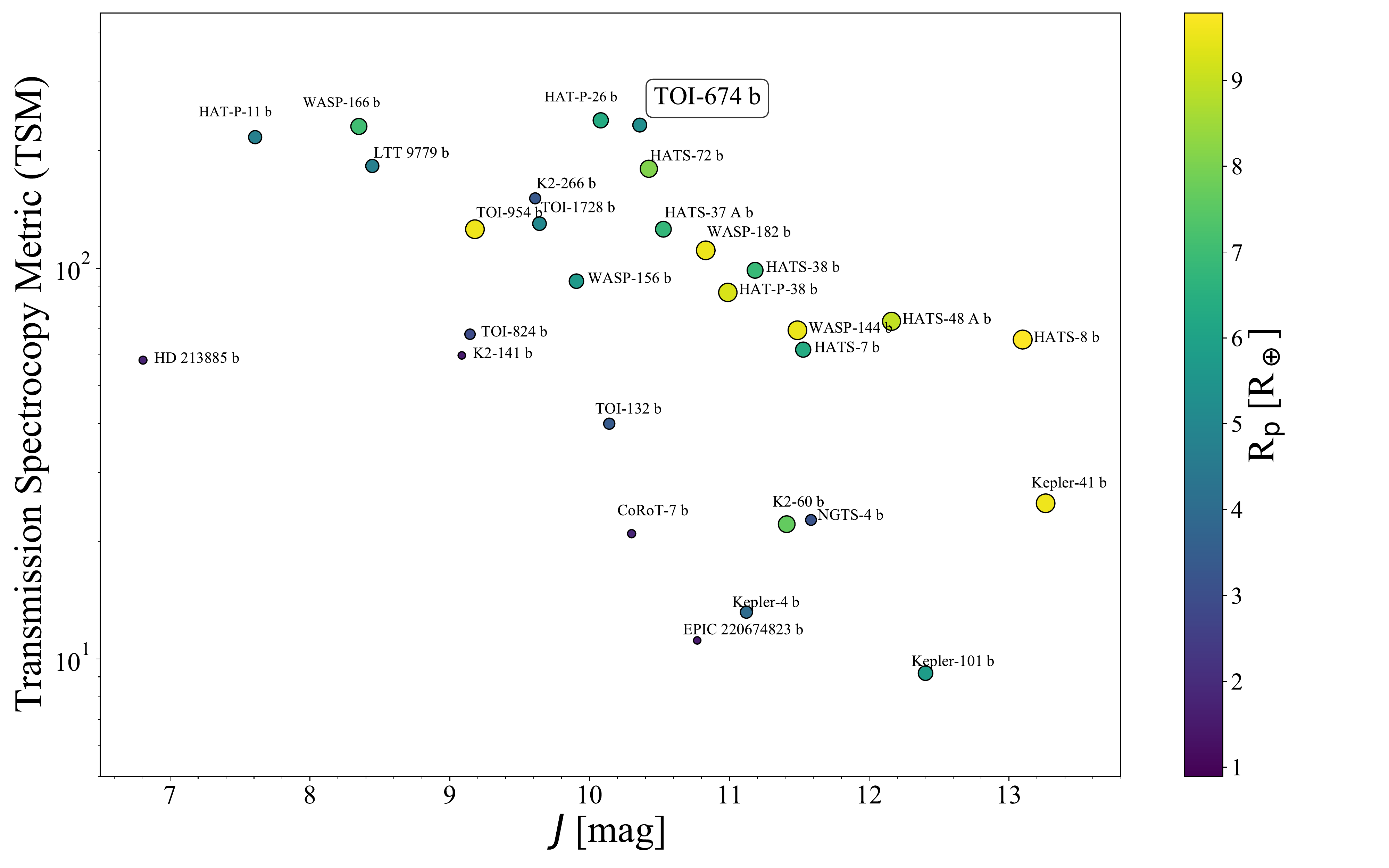}
   \caption{Apparent magnitude in \textit{J} band versus the Transmission Spectroscopy Metric (TSM) by \cite{Kempton2018} for TOI-674b and known transiting super-Earths and Neptune-sized planets inside the Neptunian Desert limits by \cite{Mazeh2016}. The size and color of each circle represents the planet radius in Earth units.}
    \label{Fig:TOI674b_TSM}
\end{figure*}

Due to the large scale height of its atmosphere and relative brightness of its host star in near infrared bands, TOI-674b is a promising target for atmospheric studies. \cite{Kempton2018} proposed a metric for the expected S/N of transmission spectroscopy with the James Webb Space Telescope (\textit{JWST}, \citealp{Gardner2006}) instrument Near Infrared Imager and Slitless Spectrograph (NIRISS) for 10 hours of observation time. Following \cite{Kempton2018}, we find a Transmission Spectroscopy Metric (TSM) of $\sim 230$ for \hbox{TOI-674b}. We compared this TSM value with those of previously discovered exoplanets that are in the Neptunian Desert (see Figure \ref{Fig:TOI674b_TSM}). TOI-674b is one of the planets belonging to the Neptunian Desert with the highest TSM factor discovered to date, making it an interesting candidate for atmospheric characterisation for \textit{JWST}.

\begin{figure*}
   \sidecaption
   \includegraphics[width=12cm]{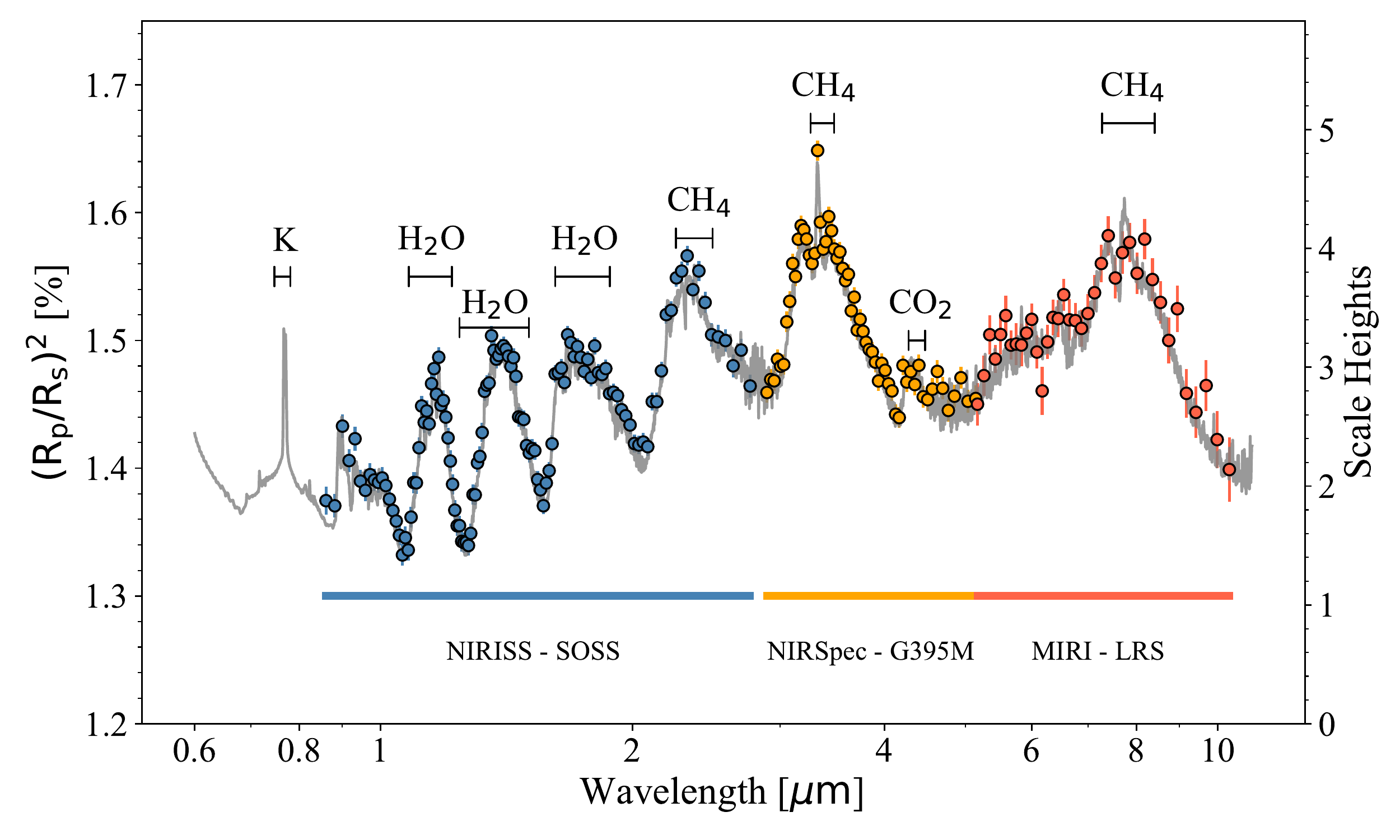} 
   \caption{Synthetic transmission spectrum model of TOI-674b (gray line) in the \textit{JWST} spectral range computed with \texttt{petitRADTRANS} (\citealp{Molliere2019}). The model assumes solar abundance and a cloud-free atmosphere with an isothermal temperature of 700 K. \texttt{PandExo} (\citeauthor{Batalha2017}) simulations of TOI-674b transmission spectrum observations made with \textit{JWST} instruments NIRISS (blue), NIRSpec (orange), and MIRI (red) are also shown.}
   \label{Fig:TOI674b_AtmosModel}
\end{figure*}

We computed a synthetic transmission spectrum of TOI-674b using \texttt{petitRADTRANS} (\citealp{Molliere2019}). The model spectrum assume solar elemental abundances and a cloud-free atmosphere with an isothermal temperature profile at a temperature of 700 K. Figure \ref{Fig:TOI674b_AtmosModel} presents model optical and infrared spectra potentially observable with \textit{JWST} for an assumed atmospheric pressure level of 0.01 bar for the continuum. The atmosphere models include molecules such as H$_2$O, CO$_2$, CO, CH$_4$, and Na and K; Rayleigh scattering from H and He is also included. 

Figure \ref{Fig:TOI674b_AtmosModel} also shows simulated observations made with \textit{JWST} of the model spectrum. The simulations were made for the \textit{JWST} instruments Near Infrared Imager and Slitless Spectrograph (NIRISS) using Single Object Slitless Spectroscopy (SOSS) mode (spectral resolution $\mathrm{R}\sim 700$), Near InfraRed Spectrograph (NIRSpec) using the medium resolution grating G395M ($\mathrm{R}\sim 1000$), and  Mid-Infrared Instrument (MIRI) using Low Resolution Spectroscopy (LRS) mode ($\mathrm{R}\sim 100$). The simulated observations cover a wavelength range of 0.8-10 $\mu$m. 

For the \textit{JWST} simulated data we used \texttt{ExoTETHyS.BOATS}\footnote{\url{https://github.com/ucl-exoplanets/ExoTETHyS}} \citep{Morello2021} to select wavelength bins with similar number of counts, in order to have comparable error bars from each point of the spectrum. Using a constant resolution, or a fixed bin width ($\Delta \lambda$), would increase the error bars and  the scatter in some spectral regions by a factor of several units, especially in the MIRI longer wavelengths. Although the total integrated S/N is independent on the choice of bins, a higher scatter per point may hinder the detection significance of broad molecular features (e.g., \citealp{Tsiaras2018}). Then, we recomputed the spectra with \texttt{PandExo} (\citealp{Batalha2017}), which fully accounts for the specific \textit{JWST} instrument modes. We noted that the spectra simulated with \texttt{ExoTETHyS.BOATS} and \texttt{PandExo} are consistent, including their error bars. The main differences appear to be related with the adopted stellar templates. Here, we show only the results obtained with \texttt{PandExo}. It is remarkable that all the various H$_2$O and CH$_4$ features can be well sampled by >10 points with order-of-magnitude smaller error bars than the amplitude of the features (for a cloud-free atmosphere). 

It should be noted that the cloud-free transmission spectrum presented here is the most optimistic scenario. Transmission spectroscopy studies done in mostly hot-Jupiters have shown that many planets posses molecular features with a lower amplitude than what the cloud-free models predict (e.g., \citealp{Sing2011}, \citealp{Deming2013}, \citealp{Kreidberg2014}). This lack of molecular traits is often explained by the presence of clouds and hazes dampening the strength of the expected features like, for example, the case of water bands in the near-infrared for hot-Jupiters (\citealp{Sing2016}). Thus, it would be preferable to establish if TOI-674b atmosphere is cloud-free with exploratory transmission spectroscopy observations before \textit{JWST} is operational. Optical low resolution observations could detect evidence for Rayleigh scattering and the broad wings of Na and K lines (e.g., \citealp{Nikolov2018}), although the detection of the latter lines would be challenging due to the relative faintness (for these type of studies) of the planet host star ($V = 14.2$ mag). In the near-infrared it would be possible to use Hubble Space Telescope (\textit{HST}) Wide Field Camera 3 (WFC3) to detect the water bands between 1.0-1.7 $\mu$m as shown by several studies (e.g., \citealp{Wakeford2013}, \citealp{Madhusudhan2014}, \citealp{Sing2016} , \citealp{Tsiaras2018}). 

Another interesting prospect for atmosphere characterisation is to search for evidence of atmospheric escape using individual lines of hydrogen (Lyman-$\alpha$, 121.5 nm) and helium (He I triplet at 1083 nm). These lines have been detected on short-period warm Neptunes orbiting around M dwarfs like for example GJ 436b (Lyman-$\alpha$, \citealp{Ehrenreich2015}) and GJ 3470b (He I, \citealp{Palle2020}). 


\section{Conclusions}
\label{Sec:Conclusions}

We report the discovery and confirm the planetary status of \hbox{TOI-674b}, a super-Neptune transiting exoplanet orbiting around a nearby M dwarf. NASA's \textit{TESS} observations led to the initial detection of the transits. Follow-up photometric observations made with several facilities and radial velocity measurements taken with the HARPS spectrograph made possible confirmation of the planetary nature of the transiting object and establishment of its mass and radius.

From a HARPS mean stellar spectrum we estimate that the spectral type of the host star is M2V and that the star has an effective an effective temperature of $\mathrm{T}_\mathrm{eff}=3514 \pm 57 $ K, and a stellar mass and radius of $\mathrm{M}_\star = 0.420\pm 0.010$ $\mathrm{M}_\odot$ and $\mathrm{R}_\star = 0.420\pm 0.013$ $\mathrm{R}_\odot$ respectively.

We analysed the data from Sector 9 and 10 two minute cadence time series observations from \textit{TESS} (PDCSAP curves) plus single transit follow-up observations from \textit{Spitzer} Space Telescope ($4.5 \; \mu \mathrm{m}$  band), and ground-based facilities: El Sauce ($\mathrm{R}_\mathrm{c}$ filter), LCOGT (Sloan-\textit{g} filter), and TRAPPIST-South telescopes ($I+z$, Sloan-\textit{z} bands). The single transit observations were detrended before performing a global joint fit of the data in order to reduce the number of free parameters and speed up the fitting procedure. The transit observations (\textit{TESS} time series plus single transit follow-up observations) and radial velocity measurements from HARPS were fitted simultaneously using a MCMC procedure that included Gaussian processes to model the systematic effects present in \textit{TESS} and RV measurements. For the joint fit we considered two cases to determine the orbital parameters of the planet: a circular orbit and a non-zero eccentricity model. 

For both fitted models (circular and eccentric model), we find that the derived planetary mass and radius agree within uncertainties. The non-zero eccentric solution presents an eccentricity value of $e = 0.10 \pm 0.05$, hence this planet could potentially join the previous known short period Neptune-sized planets with significant eccentricities. Since most of our radial velocity measurements were made during quadrature, more follow-up observations can help to further constraint the eccentricity value for this planet.

Using our circular orbit model fit we find that TOI-674b has a radius of $\mathrm{R}_\mathrm{p} = 5.25 \pm 0.17$ $\mathrm{R}_\oplus$ and a mass of $\mathrm{M}_\mathrm{p} = 23.6 \pm 3.3$ $\mathrm{M}_\oplus$, and it orbits its star with a period of $1.977143 \pm 3\times 10^{-6}$ days. We derived a mean bulk density of $\rho_\mathrm{p} = 0.91 \pm 0.15$ g cm$^{-3}$, indicating a large atmosphere/envelope. Comparing the mass and radius of TOI-674b with literature values for planets discovered orbiting around M type stars, we find that TOI-674b is one of the largest and most massive super-Neptune class planet discovered around an M-dwarf to date.

This planet is a new addition to the so-called Neptunian Desert. The Transmission Spectroscopy Metric (TSM) of TOI-674b is $\sim 240$, one of the highest TSM for Neptunian Desert planets. Thus, TOI-674b is a promising candidate for atmospheric characterisation using \textit{JWST}.

We searched for evidence of another planets in the system by measuring the transit timing variations using the data from \textit{TESS}, \textit{Spitzer}, and ground-based follow-up observations. We find no significant deviations from a linear ephemeris for the central time of the transits in a time baseline of $\sim 200$ days. TOI-674 is expected to be observed again by \textit{TESS} during its extended campaign. The observations will be done from 7 March 2021 until 2 April 2021 (corresponding to Sector 36); these new \textit{TESS} observations will help to refine TOI-674b’s orbital parameters and in the search for additional planets in the system.

\begin{acknowledgements}
N.A.-D. acknowledges the support of FONDECYT 3180063. D. D. acknowledges support from the \textit{TESS} Guest Investigator Program grant 80NSSC19K1727 and NASA Exoplanet Research Program grant 18-2XRP18\_2-0136. X. D., T. F., and G. G. acknowledge funding in the framework of the Investissements d'Avenir program (ANR-15-IDEX-02), through the funding of the ``Origin of Life'' project of the Univ. Grenoble-Alpes. G. M. has received funding from the European Union's Horizon 2020 research and innovation programme under the Marie Sk\l{}odowska-Curie grant agreement No. 895525. This paper includes data collected by the \textit{TESS} mission, which are publicly available from the Mikulski Archive for Space Telescopes (MAST). Funding for the \textit{TESS} mission is provided by NASA's Science Mission directorate. Resources supporting this work were provided by the NASA High-End Computing (HEC) Program through the NASA Advanced Supercomputing (NAS) Division at Ames Research Center for the production of the SPOC data products. This research has made use of the Exoplanet Follow-up Observation Program website, which is operated by the California Institute of Technology, under contract with the National Aeronautics and Space Administration under the Exoplanet Exploration Program. The research leading to these results has received funding from  the ARC grant for Concerted Research Actions, financed by the Wallonia-Brussels Federation. TRAPPIST is funded by the Belgian Fund for Scientific Research (Fond National de la Recherche Scientifique, FNRS) under the grant FRFC 2.5.594.09.F, with the participation of the Swiss National Science Fundation (SNF). MG and EJ are F.R.S.-FNRS Senior Research Associate. Some of the Observations in the paper made use of the High-Resolution Imaging instrument Zorro. Zorro was funded by the NASA Exoplanet Exploration Program and built at the NASA Ames Research Center by Steve B. Howell, Nic Scott, Elliott P. Horch, and Emmett Quigley. Zorro was mounted on the Gemini South telescope of the international Gemini Observatory, a program of NSF’s OIR Lab, which is managed by the Association of Universities for Research in Astronomy (AURA) under a cooperative agreement with the National Science Foundation. on behalf of the Gemini partnership: the National Science Foundation (United States), National Research Council (Canada), Agencia Nacional de Investigaci\'{o}n y Desarrollo (Chile), Ministerio de Ciencia, Tecnolog\'{i}a e Innovaci\'{o}n (Argentina), Minist\'{e}rio da Ci\^{e}ncia, Tecnologia, Inovações e Comunica\c{c}\~{o}es (Brazil), and Korea Astronomy and Space Science Institute (Republic of Korea). This work was supported by FCT - Funda\c{c}\~{a}o para a Ci\^{e}ncia e a Tecnologia through national funds and by FEDER through
COMPETE2020 - Programa Operacional Competitividade e Internacionaliza\c{c}\~{a}o by these grants: UID/FIS/04434/2019; UIDB/04434/2020; UIDP/04434/2020; PTDC/FIS-AST/32113/2017 \& POCI-01-0145-FEDER-032113; PTDC/FISAST/28953/2017 \& POCI-01-0145-FEDER-028953.

This work made use of \texttt{tpfplotter} by J. Lillo-Box (publicly available in \url{www.github.com/jlillo/tpfplotter}), which also made use of the python packages \texttt{astropy}, \texttt{lightkurve}, \texttt{matplotlib} and \texttt{numpy}. This work makes use of observations from the Las Cumbres Observatory Global Telescope Network. This work made use of \texttt{corner.py} by Daniel Foreman-Mackey (\citealp{ForemanMackey2016}).
\end{acknowledgements}

%
%

\bibliographystyle{aa}
\bibliography{references}


\begin{appendix}
\label{Sec:Appendix}

\section{Spectral energy distribution of TOI-674}

\begin{figure}
   \centering
   \includegraphics[width=\hsize]{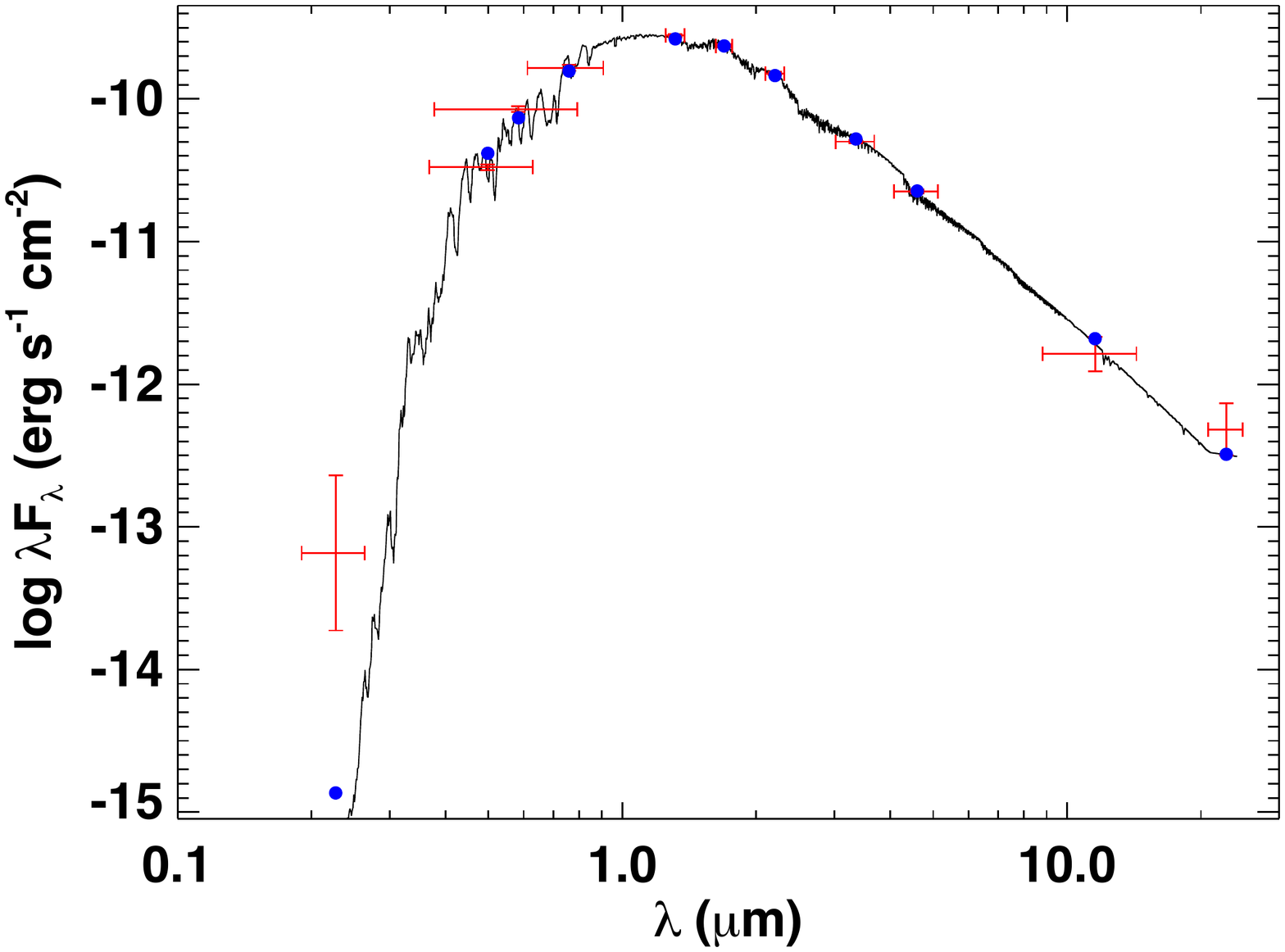}
   \caption{Spectral energy distribution of TOI-674. Red symbols represent the observed photometric measurements, where the horizontal bars represent the effective width of the passband. Blue symbols are the model fluxes from the best-fit NextGen atmosphere model (black).}
    \label{Fig:TOI674_SED}
\end{figure}

\section{Spectroscopic data measurements}

\begin{sidewaystable*}
\caption{HARPS radial velocity time series and spectroscopic activity indicators for TOI-674.}             
\label{Table:HARPS_RV}      
\centering          
\begin{tabular}{l c c c c c c c c c c c c}     
\hline\hline
BJD  [-245800] & RV  [ms$^{-1}$] & $\sigma_{\mathrm{RV}}$ [ms$^{-1}$] & $\mathrm{H}_\alpha$ & $\sigma_\mathrm{H_\alpha}$ & $\mathrm{H}_\beta$ & $\sigma_\mathrm{H_\beta}$ & $\mathrm{H}_\gamma$ & $\sigma_\mathrm{H_\gamma}$ & NaD & $\sigma_\mathrm{NaD}$ & S & $\sigma_\mathrm{S}$ \\ 
\hline                    
8627.630457 & 13481.91 & 10.85 & 0.0692 & 0.0014 & 0.0521 & 0.0037 & 0.1076 & 0.0130 & 0.0059 & 0.0021 & 0.4891 & 0.6417\\
8636.597318 & 13493.11 & 04.75 & 0.0675 & 0.0006 & 0.0495 & 0.0017 & 0.1085 & 0.0064 & 0.0076 & 0.0008 & 1.1554 & 0.3996\\
8637.628554 & 13470.93 & 06.80 & 0.0670 & 0.0008 & 0.0550 & 0.0028 & 0.1026 & 0.0106 & 0.0075 & 0.0012 & 0.7150 & 0.5353\\
8638.635832 & 13491.23 & 14.92 & 0.0668 & 0.0018 & 0.0503 & 0.0063 & 0.1001 & 0.0204 & 0.0069 & 0.0030 & 1.3946 & 0.8684\\
8639.622186 & 13476.06 & 11.65 & 0.0663 & 0.0014 & 0.0486 & 0.0050 & 0.1009 & 0.0161 & 0.0087 & 0.0022 & 0.4529 & 0.7547\\
8640.626096 & 13490.04 & 07.86 & 0.0679 & 0.0010 & 0.0452 & 0.0031 & 0.0981 & 0.0129 & 0.0081 & 0.0014 & 0.8246 & 0.6512\\
8641.611199 & 13479.48 & 10.45 & 0.0664 & 0.0013 & 0.0477 & 0.0044 & 0.1143 & 0.0192 & 0.0061 & 0.0020 & 1.1750 & 1.1082\\
8642.605629 & 13497.19 & 06.66 & 0.0677 & 0.0008 & 0.0517 & 0.0030 & 0.1134 & 0.0137 & 0.0083 & 0.0012 & 0.9768 & 0.6242\\
8643.595060 & 13483.99 & 06.48 & 0.0677 & 0.0008 & 0.0507 & 0.0026 & 0.1185 & 0.0103 & 0.0082 & 0.0011 & 0.5704 & 0.5792\\
8644.544065 & 13493.96 & 04.58 & 0.0667 & 0.0006 & 0.0524 & 0.0017 & 0.1264 & 0.0066 & 0.0088 & 0.0007 & 1.0437 & 0.4021\\
8645.519317 & 13462.41 & 08.60 & 0.0696 & 0.0011 & 0.0547 & 0.0030 & 0.1080 & 0.0099 & 0.0098 & 0.0016 & 1.0369 & 0.4539\\
8657.546337 & 13469.92 & 05.71 & 0.0690 & 0.0007 & 0.0602 & 0.0023 & 0.1071 & 0.0077 & 0.0092 & 0.0010 & 0.9712 & 0.4522\\
8660.523300 & 13490.75 & 09.88 & 0.0675 & 0.0012 & 0.0539 & 0.0037 & 0.0863 & 0.0125 & 0.0075 & 0.0019 & 1.0510 & 0.7247\\
8670.508395 & 13518.13 & 06.02 & 0.0692 & 0.0008 & 0.0500 & 0.0023 & 0.1152 & 0.0084 & 0.0072 & 0.0010 & 0.6041 & 0.3814\\
8676.515917 & 13523.37 & 09.43 & 0.0674 & 0.0012 & 0.0455 & 0.0034 & 0.1219 & 0.0121 & 0.0081 & 0.0018 & 1.0161 & 0.4858\\
8679.482254 & 13456.72 & 07.26 & 0.0680 & 0.0008 & 0.0623 & 0.0035 & 0.0900 & 0.0115 & 0.0095 & 0.0013 & 0.7379 & 0.4551\\
8682.501971 & 13504.79 & 04.83 & 0.0670 & 0.0006 & 0.0533 & 0.0018 & 0.1120 & 0.0065 & 0.0090 & 0.0008 & 0.8569 & 0.3134\\
\hline                  
\end{tabular}
\end{sidewaystable*}

\begin{figure*}
   \centering
   \includegraphics[width=\hsize]{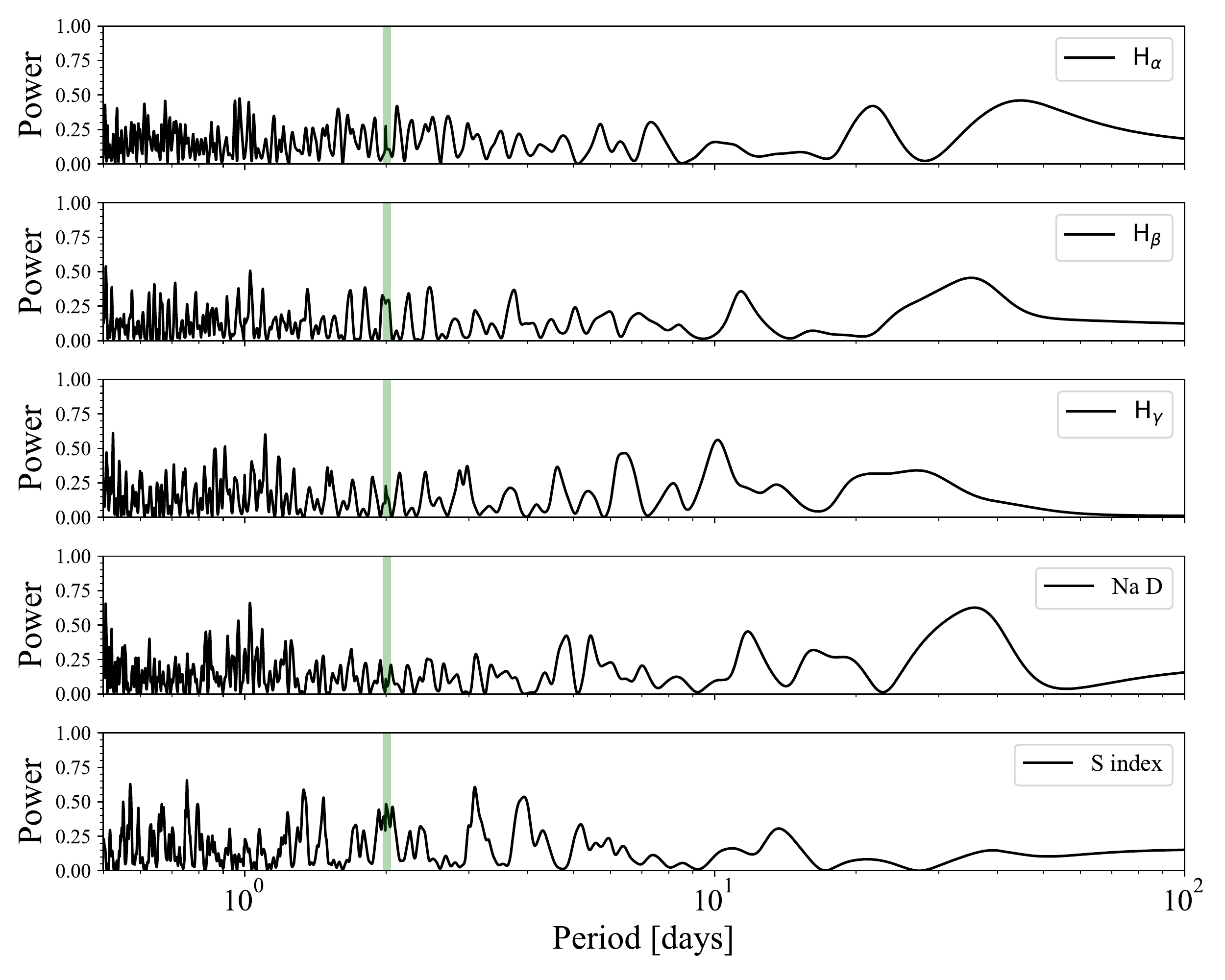}
   \caption{GLS periodograms for the activity indices obtained from 17 HARPS spectra. The period of the planet candidate is shown by the vertical green line. }
    \label{Fig:TOI674_ActivIndx_Periodogram}
\end{figure*}

\section{Light curve and radial velocity joint fit}

\begin{table*}
  \caption[]{Global fit parameters prior functions and limits; $\mathcal{U}$, $\mathcal{N}$, and $\mathcal{J}$ represent Uniform, Normal, and Jeffreys prior functions respectively.}
  \label{Table:TOI674b_ParamsPriors}
  \centering
  \begin{tabular}{lcc}
    \hline \hline
    Fitted parameter &  Ccircular orbit ($e=0$) & Non-circular orbit ($e \neq 0$) \\
    \hline

$R_\mathrm{p}/R_\star$ & $\mathcal{U}(0.05,0.35)$ & $\mathcal{U}(0.05,0.35)$ \\
$T_{\mathrm{c\;BJD}}$ [days] & $\mathcal{U}(2458640.9, 2458641.9)$ & $\mathcal{U}(2458640.9, 2458641.9)$ \\
$P$ [days] & $\mathcal{U}(1.7, 2.1)$ & $\mathcal{U}(1.7, 2.1)$ \\
$\rho_\star$ [g cm$^{-3}$] & $\mathcal{U}(5,15)$ & $\mathcal{N}(7.99,1.5)$ \\
$b = (a/R_\star) \cos(i) \left( \frac{1-e^2}{1+e\sin(\omega)} \right)$ & $\mathcal{U}(0,1)$ & $\mathcal{U}(0,1)$ \\
$\sqrt{e}\cos(\omega)$ &  & $\mathcal{U}(-1,1)$ \\
$\sqrt{e}\sin(\omega)$ &  & $\mathcal{U}(-1,1)$ \\
$\gamma_0$ [m/s] & $\mathcal{U}(-10^5,10^5)$ & $\mathcal{U}(-10^5,10^5)$ \\
$K_\mathrm{RV}$ [m/s] & $\mathcal{U}(0,110)$ & $\mathcal{U}(0,110)$ \\
$\sigma_{\mathrm{RV\; jitter}}$ [m/s] & $\mathcal{U}(0,10)$ & $\mathcal{U}(0,10)$ \\
\hline
LD coefficients & & \\
\hline
$q_1 = (u_1 + u_2)^2$ & $\mathcal{U}(0,1)$ & $\mathcal{U}(0,1)$ \\
$q_2 = 0.5u_1/(u_1+u_2)$ & $\mathcal{U}(0,1)$ & $\mathcal{U}(0,1)$ \\ 
\hline
\textit{TESS} GP parameters & & \\
\hline
$\mathrm{c}_1$ & $\mathcal{J}(10^{-4},100)$ & $\mathcal{J}(10^{-4},100)$ \\
$\tau_1$ [days] & $\mathcal{J}(10^{-4},200)$ & $\mathcal{J}(10^{-4},200)$ \\
\hline
HARPS GP parameters & & \\
\hline
$\mathrm{c}_2$ [m/s] & $\mathcal{U}(0,100)$ & $\mathcal{U}(0,100)$ \\
$\tau_2$ [days] & $\mathcal{U}(10^{-3},80)$ & $\mathcal{U}(10^{-3},80)$ \\
\hline
\end{tabular}
\tablefoot{We also imposed $e<1$ for the non-circular orbit fit.}
\end{table*}

\begin{figure*}
   \centering
   \includegraphics[width=\textwidth]{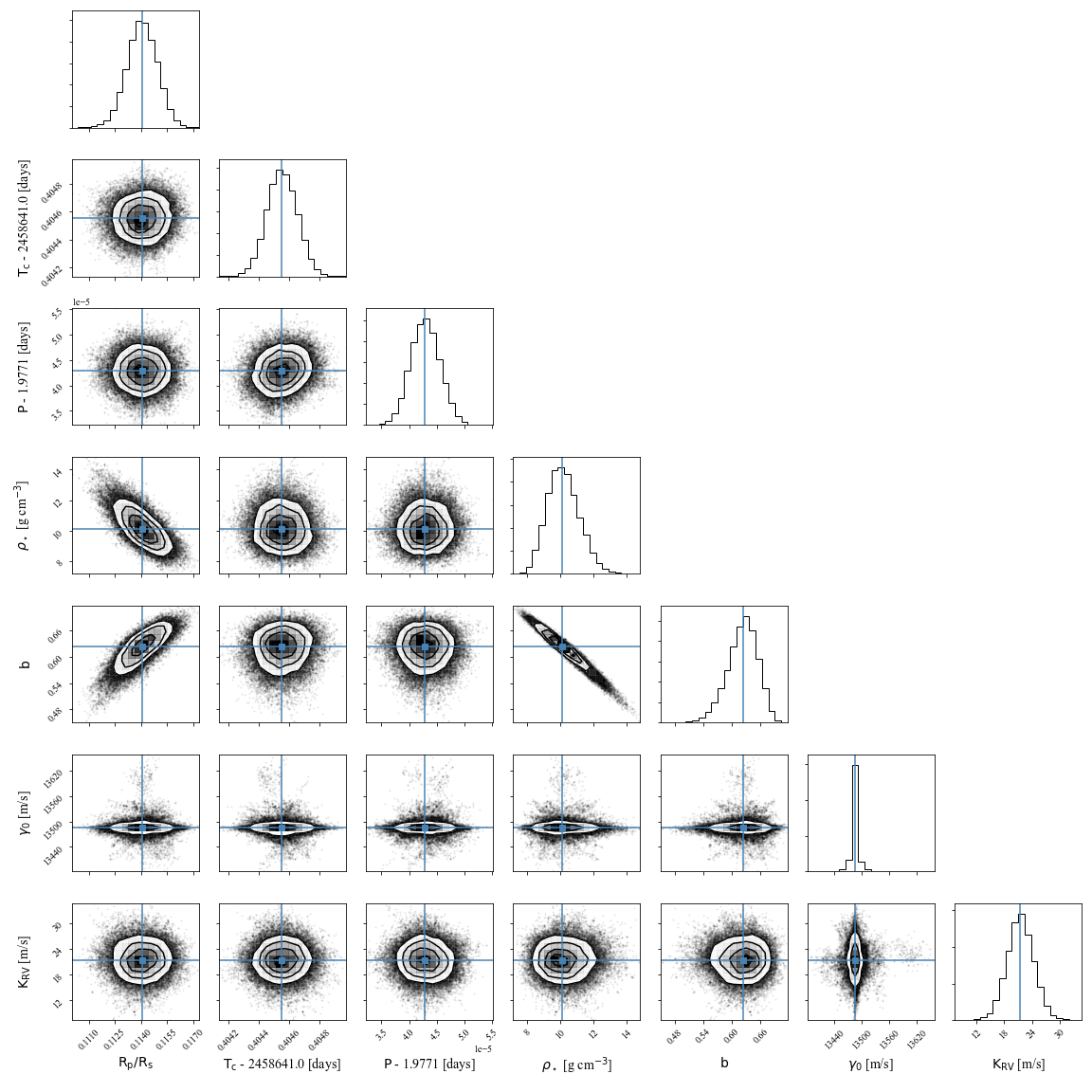}
   \caption{Correlation plot for the orbital parameters fitted using a circular orbit model. Limb darkening coefficients and systematic effect parameters were left out intentionally for easy viewing. The blue lines mark the median values of the parameters.}
    \label{Fig:Fit_Circular_CornerPlot}
\end{figure*}

\begin{figure*}
   \centering
   \includegraphics[width=\textwidth]{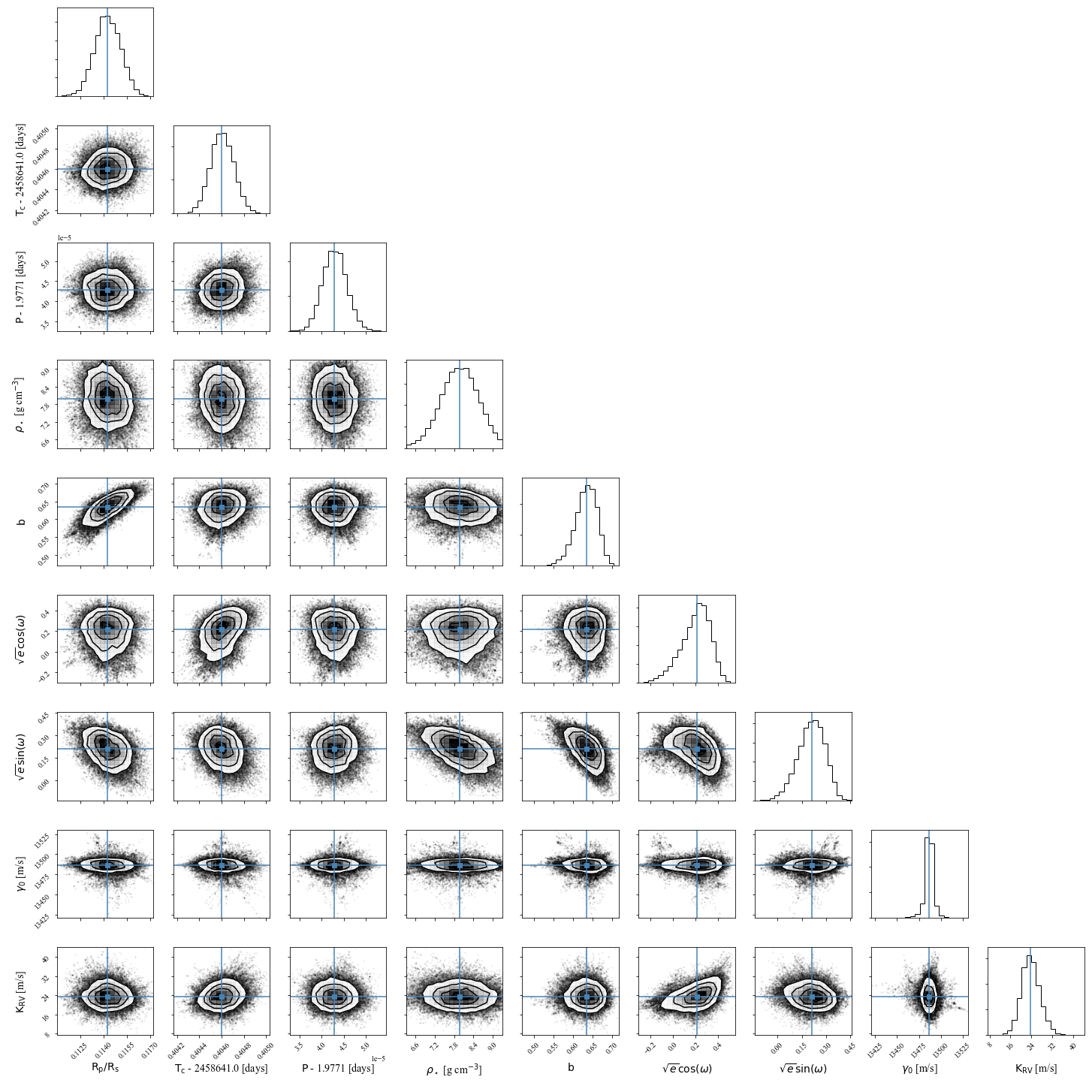}
   \caption{Correlation plot for the orbital parameters fitted using a non-zero eccentricity orbit model. Limb darkening coefficients and systematic effect parameters were left out intentionally for easy viewing. The blue lines mark the median values of the parameters.}
    \label{Fig:Fit_Eccentric_CornerPlot}
\end{figure*}


\begin{figure*}
   \centering
   \includegraphics[width=\textwidth]{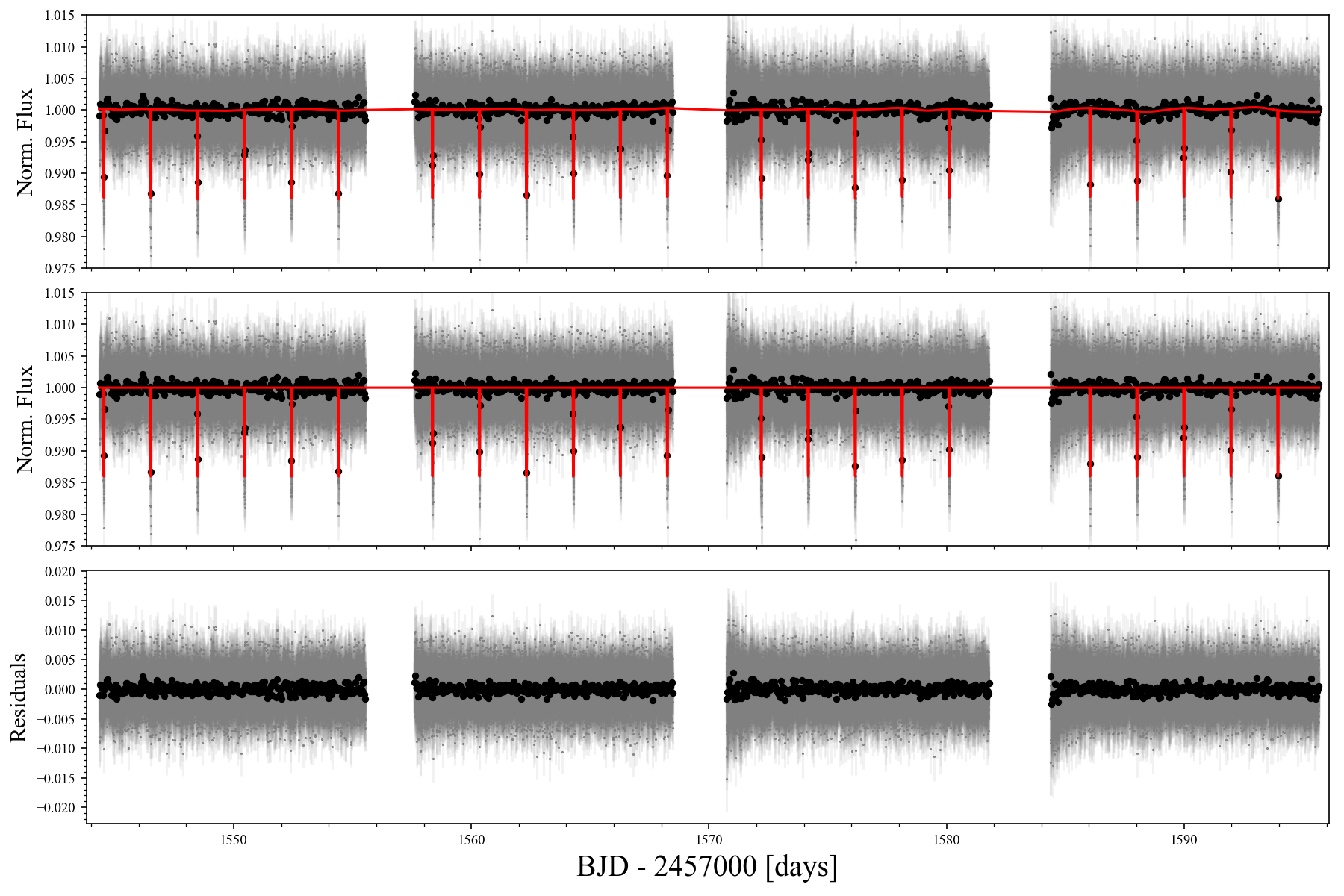}
   \caption{\textit{TESS} Sector 9 and Sector 10 two minute cadence light curves of TOI-674. \textit{Top panel}: \textit{TESS} PDCSAP flux time series and best circular orbit model from the joint fit. \textit{Middle panel}: \textit{TESS} flux and transit model after removing stellar variability.\textit{Bottom panel}: residuals of the fit.}
    \label{Fig:TESS_LC_S9_S10}
\end{figure*}

\end{appendix}

\end{document}